\documentclass[sigconf]{acmart}
\AtBeginDocument{%
  }

\setcopyright{acmlicensed}
\copyrightyear{2026}
\acmYear{2026}
\acmDOI{XXXXXXX.XXXXXXX}
\acmConference[UMAP '26]{June 08--11, 2026}{Gothenburg, Sweden}

\usepackage{subcaption}
\usepackage{graphicx}
\usepackage[utf8]{inputenc}
\usepackage{textcomp}
\usepackage{newunicodechar}
\newunicodechar{−}{-}
\begin{document}

\title[Visual vs. Textual Explanations in an ERS]{Visual or Textual: Effects of Explanation Format and Personal Characteristics on the Perception of Explanations in an Educational Recommender System}
\author{Qurat Ul Ain}
\email{qurat.ain@stud.uni-due.de}
\affiliation{%
  \institution{University of Duisburg-Essen}
  \city{Duisburg}
  \country{Germany}
}
\author{Mohamed Amine Chatti}
\email{mohamed.chatti@uni-due.de}
\affiliation{%
  \institution{University of Duisburg-Essen}
  \city{Duisburg}
  \country{Germany}
}
\author{Nasim Yazdian Varjani}
\email{nasim.yazdian@rwth-aachen.de}
\affiliation{%
  \institution{RWTH Aachen University}
  \city{Aachen}
  \country{Germany}
}
\author{Farah Kamal}
\email{farah.kamal@stud.uni-due.de}
\affiliation{%
  \institution{University of Duisburg-Essen}
  \city{Duisburg}
  \country{Germany}
}
\author{Astrid Rosenthal-von der Pütten}
\email{arvdp@itec.rwth-aachen.de}
\affiliation{%
  \institution{RWTH Aachen University}
  \city{Aachen}
  \country{Germany}
}
\renewcommand{\shortauthors}{Ain et al.}
\begin{abstract}
Explanations are central to improving transparency, trust, and user satisfaction in recommender systems (RS), yet it remains unclear how different explanation formats (visual vs. textual) are suited to users with different personal characteristics (PCs).
To this end, we report a within-subject user study (n=54) comparing visual and textual explanations and examine how explanation format and PCs jointly influence perceived control, transparency, trust, and satisfaction in an educational recommender system (ERS). Using robust mixed-effects models, we analyze the moderating effects of a wide range of PCs, including Big Five traits, need for cognition, decision making style, visualization familiarity, and technical expertise.
Our results show that a well-designed visual, simple, interactive, selective, easy to understand visualization that clearly and intuitively communicates how user preferences are linked to recommendations, fosters perceived control, transparency, appropriate trust, and satisfaction in the ERS for most users, independent of their PCs. Moreover, we derive a set of guidelines to support
the effective design of explanations in ERSs.  
\end{abstract}

\maketitle
\section{Introduction}
Explanations have become an integral component of modern recommender systems (RSs) \cite{zhang2020explainable,chatti2024visualization}. Explainability refers to providing human-understandable information about why a particular item was recommended and how the system works internally \cite{tintarev2007survey}. Explanations aim to improve users perceptions of trust, transparency, scrutability, effectiveness,
efficiency, persuasiveness, and satisfaction with the system, which in turn can lead to better acceptance of recommendations \cite{balog2020measuring, dominguez2019effect, ooge2022explaining, millecamp2019explain, guesmi2024interactive, al2025f}. Explanations are typically presented to the users either using visualizations (i.e., visual), or in natural language description (i.e., textual), referred to as explanation format \cite{ain2022multi}. 
Prior research suggests that users’ perceptions of RS and its explanations depend on how explanations are presented and are further shaped by individual differences \cite{millecamp2021textualvisual,kouki2019personalized,chatti2022more,hernandez2021explaining}. 
While explanations have been evaluated in terms of various explanation styles \cite{kouki2019personalized} and varying level of details \cite{guesmi2024interactive}, empirical evidence directly comparing explanation formats remains limited \cite{kouki2019personalized, millecamp2021textualvisual}. As a result, it is still unclear whether users perceive visual or textual explanations as more effective.
Since explanations aim to help users understand the recommended items and make accurate decisions \cite{lu2023user}, their role is particularly important in educational settings, where decision making can directly influence learning outcomes. Although explanations have recently been investigated in the context of educational recommender systems (ERS) \cite{ooge2022explaining}, to the best of our knowledge, the comparative effects of explanation formats on users’ perceptions of ERS remain underexplored. In parallel, despite growing interest in incorporating personality into RS \cite{tintarev2017presenting}, only a limited number of studies have examined how personal characteristics (PCs) influence users’ perceptions of explanations in RS \cite{millecamp2019explain, kouki2019personalized, naveed2018argumentation, guesmi2024interactive,chatti2022more,hernandez2021explaining}. Furthermore, the impact of PCs on the perception of explanations in ERSs has not yet been studied. 

To bridge these gaps, in this work we systematically compare visual vs. textual explanations within an ERS and investigate how explanation format influences users’ perceptions of the system. Specifically, we examine the effects of explanation format on perceived control, transparency, trust, and satisfaction. In addition, we analyze how PCs, including Big Five traits, Need for Cognition (NFC), Decision Making Style (DMS), Visualization Familiarity (VF) and Technical Expertise (TE) moderate users perceptions of explanation formats. 
The following research questions guide our investigation:

\textbf{RQ1.} How does explanation format (visual vs. textual) impact users' perceptions of control, transparency, trust, and satisfaction with an ERS?

\textbf{RQ2.} How do personal characteristics influence users’ perceptions of visual and textual explanations in terms of perceived control, transparency, trust, and satisfaction?

To answer these RQs, we conducted a within-subject user study (n=54) using a mixed-method evaluation approach. Our results show that while textual and visual explanations supported comparable levels of perceived control and transparency, visual explanations significantly fostered higher trust and satisfaction. Furthermore, 
users with higher Agreeableness perceived more control and users with higher Conscientiousness perceived more trust, and satisfaction with visual explanations. Whereas, users with low intuitive DMS significantly trust visual explanations more. 
Overall, visual explanations yielded higher perceived control, transparency, trust, and satisfaction, with these benefits remaining largely consistent across users regardless of PCs.
\section{Background and Related Work}
\subsection{Explanation Format}
The explanation format refers to the way in which explanations are presented to the users \cite{ain2022multi}. Explanations can be conveyed through visual representations such as images, graphs, or charts (i.e., visual) \cite{chatti2024visualization} or through natural language descriptions (i.e., textual) \cite{zhang2020explainable,al2025f}. Visual explanations typically rely on graphical representations to communicate explanatory information. Common visualization techniques include bar charts, pie charts, histograms, tag clouds, saliency maps, scatter plots, line charts, and Venn diagrams \cite{gedikli2014should, guesmi2024interactive, vig2009tagsplanations, adadi2018peeking, wang2019designing, chatti2022more, guesmi2021input, tsai2019explaining, yang2020visual}. In addition to charts, images have also been used as visual explanations \cite{lin2019explainable, chen2019personalized}.
For more details, we refer the interested readers to a recent comprehensive survey of visually explainable recommendation \cite{chatti2024visualization}. 
The most frequently used approach for generating textual explanations is template-based using predefined sentence templates that are populated based on the underlying recommendation algorithm \cite{nunes2017systematic}. These templates may incorporate various factors, such as input parameters and item features \cite{millecamp2021textualvisual, kim2018textual, lu2023user}, similarity to other items \cite{kunkel2019let, kouki2019personalized}, usage context \cite{sato2019context}, or similarity to other users \cite{kouki2019personalized, lu2023user}. 
Another method of generating textual explanations without templates is natural language explanation that generates explanation sentences automatically using NLP and machine learning techniques \cite{chang2016crowd, musto2016explod, zhao2019personalized, zhao2018you, costa2018automatic, musto2019justifying, sun2021unsupervised, xian2020cafe}, 
and more recently, employing LLMs \cite{hada2021rexplug, yang2024fine}.
Both explanation formats have proved to positively influence users’ perceptions of RSs, including transparency \cite{ooge2022explaining, zhao2019users, gedikli2014should}, trust \cite{yang2020visual,ooge2022explaining, millecamp2019explain}, and satisfaction \cite{musto2019justifying, gedikli2014should, millecamp2019explain}. Moreover, interactive explanations have proven to positively impact users perceived control \cite{guesmi2021input,guesmi2024interactive}. 

Prior research on explainable RS has mainly focused on comparing recommendations with and without explanations \cite{millecamp2019explain}, different explanation styles \cite{kouki2019personalized, herlocker2000explaining}, or varying level of details \cite{guesmi2024interactive}. However, although explanation format also influences user experience \cite{kouki2017user}, direct comparisons between visual and textual explanations remain rare. Notable exceptions include Kouki et al. \cite{kouki2019personalized}, who found textual explanations to be more persuasive than visual ones, and Millecamp et al. \cite{millecamp2021textualvisual}, who reported that although users preferred visual explanations, lay users performed better with textual explanations.
These findings highlight the need for further systematic comparisons of explanation formats to better understand how they influence users’ perceptions of explanations.
Moreover, to the best of our knowledge, the comparative effects of explanation formats have not yet been systematically investigated in the context of ERSs. To address this gap, in this work we compare visual vs. textual explanations together in an ERS to investigate which format is best to convey explanatory information, to meet different explanation aims.  
\subsection{Personal Characteristics}
Recent research has demonstrated that PCs influence how users perceive explanations in RSs 
\cite{chatti2022more, hernandez2021explaining, kouki2019personalized, millecamp2021textualvisual, millecamp2019explain, millecamp2020s, martijn2022knowing}.
For instance, \citet{millecamp2019explain} investigated the role of Big Five traits and NFC in the presence versus absence of explanations. 
Extending this line of work, \citet{martijn2022knowing} investigated perception of explanation of users with low vs. high NFC, musical sophistication, and openness.
Furthermore, \citet{kouki2019personalized} examined how Big Five traits and VF relate to user preferences for explanation styles. 
Other PCs have also been explored in literature including DMS \cite{naveed2018argumentation}, user expertise \cite{millecamp2021textualvisual}, personal innovativeness, trust propensity, and domain knowledge \cite{chatti2022more,guesmi2022explaining}.
Overall, prior work clearly indicates that PCs play an important role in shaping users’ perceptions of explanations and should be considered when designing explainable RSs. However, the effects of PCs in comparing different explanation formats remain underexplored \cite{millecamp2021textualvisual,kouki2019personalized}. 
\section{Visual and Textual Explanations in CourseMapper}
In our ERS in CourseMapper, that recommends YouTube videos and Wikipedia articles to learners, we provide both visual and textual explanations to help learners understand why an item is recommended based on their selected inputs.
The videos and articles are recommended based on the similarity score between keyphrases extracted from these resources and concepts that the user marked as "Did Not Understand" when interacting with a learning material (referred to as DNU concepts).
Both visual and textual explanations draw upon this information.
\subsection{Visual Explanation Design}
The design of visual explanations in CourseMapper is adapted from our earlier research on interactive explanations developed in RIMA \cite{guesmi2024interactive}. These explanations were systematically designed using a Human-Centered Design (HCD) approach and popular categorization of intelligibility types \cite{lim2009assessing} to come up with "what", "why", "how", and "what-if" explanation with different levels of detail. Evaluating those explanations in RIMA \cite{guesmi2024interactive, chatti2022more} revealed that the users found "what" and "why" explanations with abstract and intermediate level of detail the most understandable and effective. Whereas, most of the users
found "how" explanation, either too technical or unnecessary, so we decided not to have it in CourseMapper. 

The "what" explanation in CourseMapper revealing to the users what does the system know about them, displays the user input (DNUs), as chips with different colors at the top of the recommendations (Figure \ref{fig:visual}-a(a)). 
The "why" explanation 
is further divided into two levels of detail, namely abstract and detailed. In the "Why (abstract)" explanation, with each recommended video and article, a cosine similarity 
score between the recommended item and the user input concepts is displayed at the top right corner of each recommendation (Figure \ref{fig:visual}-a(c)). Moreover, a color band on the left 
side of the recommended item indicates the similarity score between the 
current item and user's each input concept (Figure \ref{fig:visual}-(b)). Furthermore, the keyphrases in the description/abstract of the recommended video/article are 
highlighted in the color of the most similar input concept (Figure \ref{fig:visual}-a(d)). Hovering over a colored keyphrase shows its similarity score with the most similar input concept.
Clicking on a keyphrase, opens a pop-up containing a bar chart that shows the similarity scores between the selected keyphrase and the user's top three most similar input concepts (Figure \ref{fig:visual}-a(e)).
Furthermore, the "Why (detailed)" explanation is 
provided using the 'Why button' (Figure \ref{fig:visual}-a(f)) which displays an interactive, colored word cloud containing all the
keyphrases extracted from the description/abstract of the recommended video/article, allowing users to see all of them
at a glance (Figure \ref{fig:visual}-a(g)). 
Hovering over any keyphrase in the word cloud, shows a
colored bar chart on the right side of the word cloud and updates dynamically as a
different keyphrase is hovered over. It represents the similarity score between the
hovered keyphrase and user’s top five inputs. 
\subsection{Textual Explanation Design}
\begin{figure*}[h]
    \centering
    \includegraphics[width=0.9\linewidth]{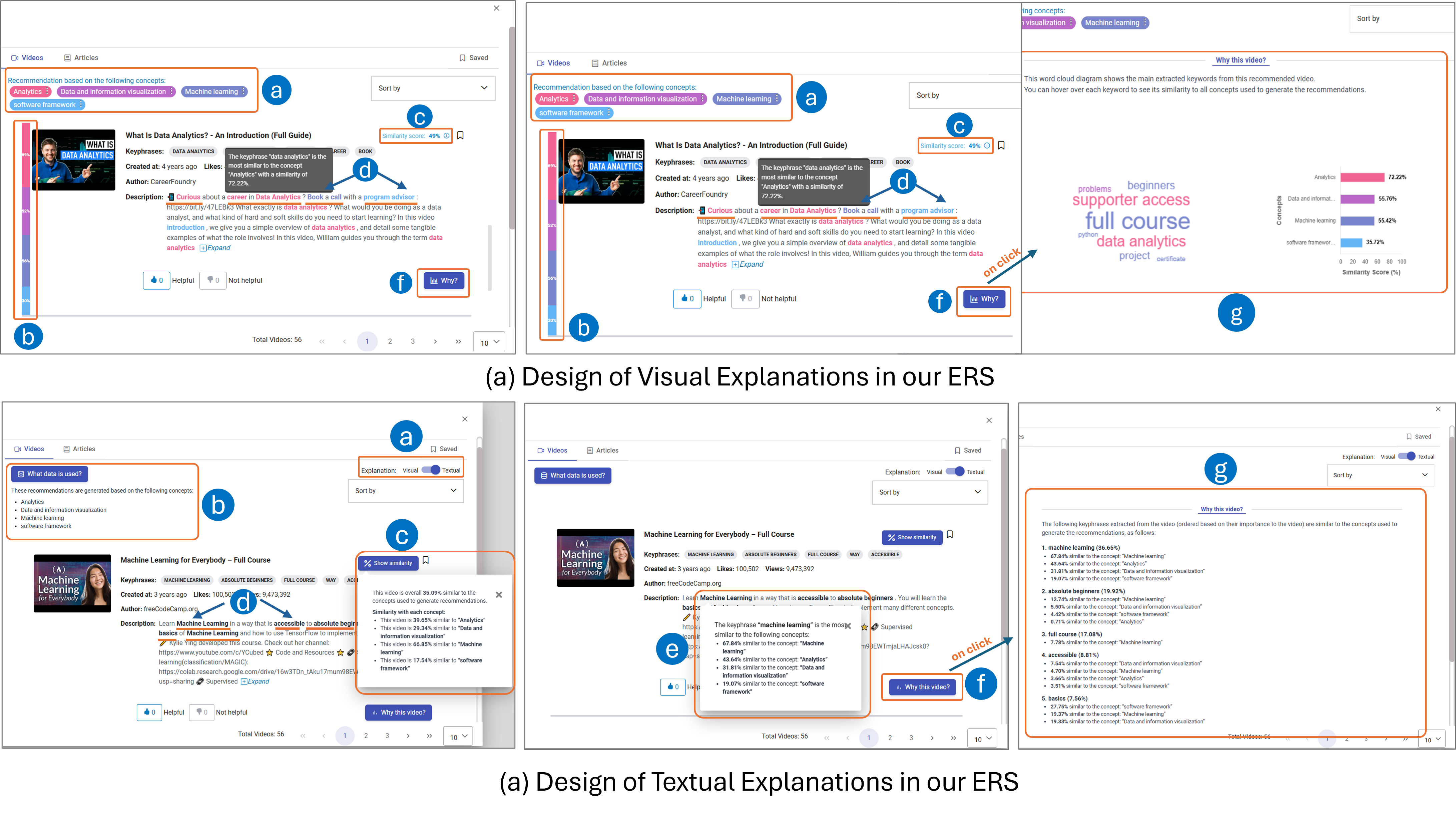}
    \Description{visual all together}
  \caption{Visual and Textual explanations in CourseMapper}
  \label{fig:visual}
\end{figure*}

For textual explanations, we translated the exact information and level of detail used in visual explanations to textual format for each intelligibility type. To ensure this one to one mapping from visual to textual format, the suitable textual explanation method was template-based explanation. This design choice ensured that both explanation formats conveyed the same information, allowing for a fair and controlled comparison between visual and textual explanations. A toggle button was used to switch between visual and textual explanation UIs (Figure \ref{fig:visual}-b(a)). 
For the "what" explanation, we display a (template-based) sentence to list down users' input concepts (Figure \ref{fig:visual}-b(b)). 
In the "why (abstract)" explanation, with each recommendation, a 'Show Similarity' button is displayed which shows the cosine similarity score between the recommended item and user's each input concept (Figure \ref{fig:visual}-b(c)).
Furthermore, the keyphrases in the description/abstract of the recommended video/article are highlighted in bold (Figure \ref{fig:visual}-b(d)). Hovering over a keyphrase, displays the similarity score in percentage between that keyphrase and the most similar input concept. Furthermore, clicking on a keyword, opens a pop-up containing textual information that shows the similarity scores between the selected keyphrase and the user's top three most similar inputs (Figure \ref{fig:visual}-b(e)).
"Why (detailed)" explanation is provided using the 'Why button' (Figure \ref{fig:visual}-b(f)). When clicked, it shows all the keyphrases extracted from the description/abstract of the recommended video/article and their similarity score with user's top five input concepts (Figure \ref{fig:visual}-b(g)), following the template for all keyphrases from i to n as:
"The following keyphrases extracted from the video/article (ordered based on their relevance to the video) are similar to the concepts used to generate recommendations as follows: 

1 to n): [$Keyphrase_i..._n$] ([percentage similarity with the video])

     \textbullet\ [similarity score \%] similar to the concept "[$C_1$]"..."[$C_5$]".
\section{User Study}
We conducted a within-subject user study, where each participant examined both visual and textual explanations. The study design and procedure was approved by the Ethics Committee of the University \textit{[Blinded Name]}.
The required sample size (n=54) was determined a priori through power analysis (effect size f = 0.25, power = 0.95) using G*Power \cite{faul2007g}, ensuring sufficient statistical power to detect medium-sized effects. The study was conducted online, participation was voluntary and took approximately one hour on average. Every participant received a monetary compensation for their participation. 
\subsection{Participants}
Participants were recruited through advertisements distributed at universities and academic networks in Germany, Pakistan, and Iran (based on authors’ institutional and professional contacts). Participants were required to be at least 18 years old and proficient in English.
A total of 54 participants (29M, 25F), aged between 19 to 60 years (M = 29.0, SD = 8.02), completed the study. 
\subsection{Measures}
To assess users’ perceptions of the RS and to capture relevant individual differences, we employed a set of standardized questionnaire measures assessing our dependent variables and PCs.
\subsubsection{Dependent variables:}
\textit{Perceived control} in RS measures if users felt in control in their interaction
with the recommender \cite{pu2011user}. It was measured on a 5-point Likert scale using a self-created item (adapted from \cite{pu2011user}): \textit{"I feel in control of the level of information/details of the explanations provided in the system"}.
\textit{Transparency} determines whether or not a system allows users to understand its inner logic, i.e. why a particular item is
recommended to them \cite{pu2011user}. 
We measured transparency 
using 12 items adopted from \cite{hellmann2022development}, on a 5-point Likert scale. 
\textit{Trust} measures an individual's willingness to depend on a specific technology \cite{mcknight2009trust}. It was evaluated using 8 items adopted from \cite{mcknight2009trust} on a 7-point Likert scale. 
\textit{Satisfaction} refers to the overall user experience with the RS, including whether users find it useful, beneficial, and worth recommending to others \cite{knijnenburg2012explaining}. It was measured using 7 items adapted from \cite{knijnenburg2012explaining} on a 5-point Likert scale.
\subsubsection{Personal Characteristics (PCs):}
To investigate individual differences in users’ responses to textual and visual explanations, we collected several PCs, measured using Likert scales in accordance with their original questionnaire specifications.

\textit{Big Five Traits}: refer to the five basic dimensions
of personality including Openness, Conscientiousness, Extraversion, Agreeableness, and Neuroticism. These were measured using the questionnaire by \citet{gosling2003very} which is both brief
and highly reliable. Given prior evidence that Big Five traits significantly affect users’ perception of explanations \cite{millecamp2020s, kouki2019personalized}, we aim to further investigate their influence on explanation formats as well.

\textit{Need for Cognition (NFC)}: refers to the tendency
for an individual to engage in and enjoy effortful cognitive activities \cite{haugtvedt1992need}. 
Previous studies show that NFC affects how users process and benefit from explanations \cite{millecamp2019explain,chatti2022more}.
Given the interactive and selective nature of our visual and textual explanations, NFC is a relevant factor for further investigation. We measure NFC using the NCS-6 scale \cite{lins2020very}.

\textit{Decision Making Style (DMS)}: captures how individuals typically process information when making decisions, with rational styles emphasizing analytical reasoning and intuitive styles relying on experience-based judgments \cite{hamilton2016development}. Prior work in the RS domain indicates that DMS influence users’ perceptions of explanations \cite{naveed2018argumentation, hernandez2021explaining}. Given that explanations aim to support decision making, DMS represents a relevant PC, motivating its inclusion in our study, measured using the 10-item Decision Styles Scale \cite{hamilton2016development}.

\textit{Visualization Familiarity (VF)}: refers to the extent to which users have experience with analyzing and
graphing data visualizations. A higher visualization familiarity has been found to positively influence users perception of visual explanations \cite{kouki2019personalized, chatti2022more}, which motivated us to further investigate this characteristic in our study.
We measured VF using 4 items adapted from \cite{kouki2019personalized}.

\textit{Technical Expertise (TE)}: captures users’ perceived knowledge of recommender systems and their ability to comprehend how recommendations are generated. TE was chosen as a PC in this study because it directly influences how different explanation modalities are understood and used by different users \cite{chatti2022more, millecamp2021textualvisual}.
TE was measured using 2 items adapted from \cite{kunkel2021identifying}.
\subsection{Study Procedure}
The study was conducted through one-to-one online sessions using Zoom. Each session began with an introduction to the study goals and procedure, followed by a pre-study questionnaire collecting demographic information and PCs. Participants then watched a short demonstration video introducing our ERS and its interface.
Participants were randomly assigned to one of two explanation conditions (textual or visual) and given screen control to interact with the system. They completed two tasks while following a think-aloud protocol. In the first task, participants enrolled in a course of their choice, selected a PDF learning material, and marked concepts they did not understand while reading, which were then used to generate YouTube video recommendations. In the second task, participants explored the explanations corresponding to their assigned condition and reflected on their experience.
After completing the tasks, participants filled out a post-study questionnaire evaluating perceived control, transparency, trust, and satisfaction. Participants then repeated the second task using the alternative explanation format, followed by the same questionnaire. Finally, participants answered open-ended questions about their experiences and preferences regarding visual and textual explanations and their impact on perceived control, transparency, trust, and satisfaction.
\subsection{Data Analysis}
Composite scores for all DVs and PCs were calculated and quantitative analyses were conducted in R. For RQ1, within-subject differences between explanations were analyzed using Wilcoxon signed-rank tests (with Holm adjustment) due to non-normality (Shapiro–Wilk, all p < .001) \cite{howell1992statistical}, and effect sizes were reported. For RQ2, moderation effects were examined using robust mixed-effects models. Significant interactions were followed up using estimated marginal means (EMMs) with 95\% confidence intervals, and robustness was confirmed using false discovery rate (FDR) correction. Qualitative responses were analyzed using thematic analysis \cite{braun2021thematic}.
  
  
\section{Results}
\subsection{Visual vs. Textual Explanations}
Participants’ perceptions of the ERS differed across explanation formats for some, but not all, evaluated dimensions. Table \ref{tab:rq1_results} summarizes the descriptive statistics. 
\begin{table*}
  \caption{Descriptive and inferential statistics for visual and textual explanations.}
  \label{tab:rq1_results}
  \centering
  \begin{tabular}{lcccccc}
    \toprule
    \textbf{Dependent Variable} &
    \textbf{Textual (M $\pm$ SD)} &
    \textbf{Visual (M $\pm$ SD)} &
    \textbf{$V$} &
    \textbf{$p$-value} &
    \textbf{Effect size ($r$)} &
    \textbf{Interpretation} \\
    \midrule
    Perceived Control        & 3.85 $\pm$ 0.98 & \textbf{3.98 $\pm$ 0.88} & 240.5 & 0.378 & 0.08 (small)     & No difference \\
    Transparency   & 3.85 $\pm$ 0.67 & \textbf{3.94 $\pm$ 0.67} & 666.0 & 0.426 & 0.10 (small)     & No difference \\
     Trust          & 5.28 $\pm$ 1.33 & \textbf{5.64 $\pm$ 1.11} & 756.0 & \textbf{0.007} & 0.36 (moderate)  & Visual $>$ Textual \\
    Satisfaction   & 4.01 $\pm$ 0.69 & \textbf{4.21 $\pm$ 0.59} & 540.5 & \textbf{0.014} & 0.35 (moderate)  & Visual $>$ Textual \\
    \bottomrule
  \end{tabular}
\end{table*}
\textit{Perceived control} did not differ significantly between explanation formats. Control ratings were comparable for textual (M = 3.85, SD = 0.98) and visual (M = 3.98, SD = 0.88) explanations, and the Wilcoxon signed-rank test indicated no statistically significant difference. 
Similarly, no significant difference was observed for perceived \textit{transparency}. Participants reported comparable transparency levels for textual (M = 3.85, SD = 0.67) and visual explanations (M = 3.94, SD = 0.67). 
In contrast, explanation format significantly affected perceived \textit{trust}, with higher ratings for visual explanations (M = 5.64, SD = 1.11) than for textual explanations (M = 5.28, SD = 1.33; V = 756, adjusted p = .007, r = .36), indicating a moderate effect. Similarly, \textit{satisfaction} was higher for visual explanations (M = 4.21, SD = 0.59) than for textual explanations (M = 4.01, SD = 0.69). This difference was statistically significant, with a moderate effect size (V = 540.5, adjusted p = .014, r = .35). Figure \ref{fig:rq1-4subfigs} summarizes these results.
\begin{figure}[h]
    \centering
    \includegraphics[scale=0.42]{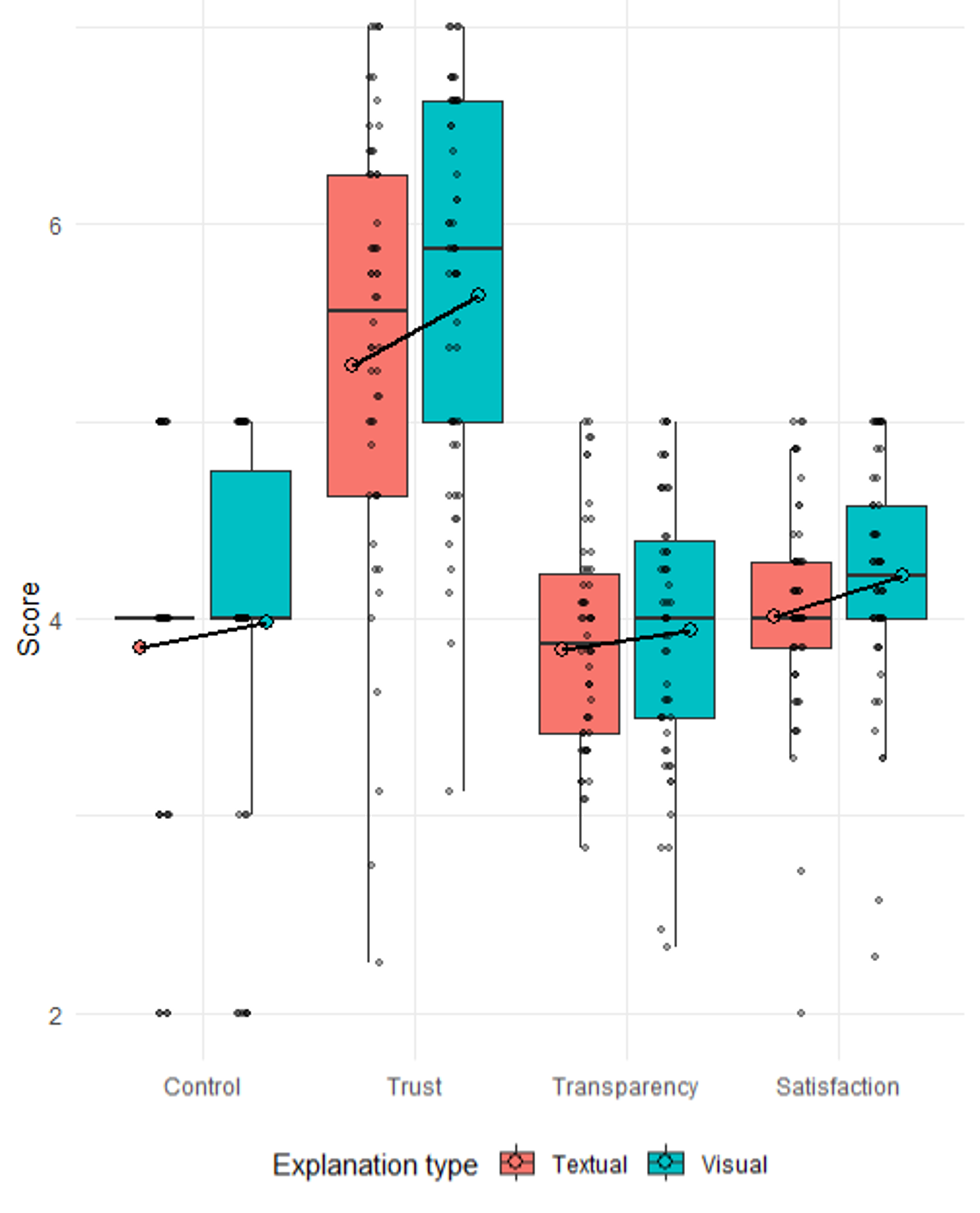}
    \Description{Boxplots all together}
  \caption{Textual vs. visual explanations across all measures}
  \label{fig:rq1-allboxes}
\end{figure}
\subsection{Impact of Personal Characteristics}
Following RQ1, we examined whether presentation order influenced users’ perceptions and whether PCs moderated the effects of explanation format. We fitted robust linear mixed-effects models of the form:
\begin{equation}
\text{DV} \sim \text{ExplanationType} \times \text{Order} 
+ \text{ExplanationType} \times \text{PC} 
+ (1 \mid \text{Participant}),
\end{equation}
Where, DV denotes dependent variables, ExplanationType (visual vs. textual), presentation order (textual first or visual first), and PC (and interactions) as fixed effects and participant as a random intercept, where moderators were mean-
centered.
The main effects of explanation format replicated the RQ1 results and are therefore not discussed further. No significant effects of presentation order or its interaction with explanation format were observed for any DV, indicating that order did not confound the results. For perceived control, a non-significant trend suggested higher ratings for visual explanations when presented at second order. Overall, presentation order did not meaningfully affect users’ evaluations, allowing moderation analyses to focus on PCs.
\begin{figure}[h]
  \centering
  \begin{subfigure}[t]{0.499\columnwidth}
    \centering
    \includegraphics[width=\linewidth]{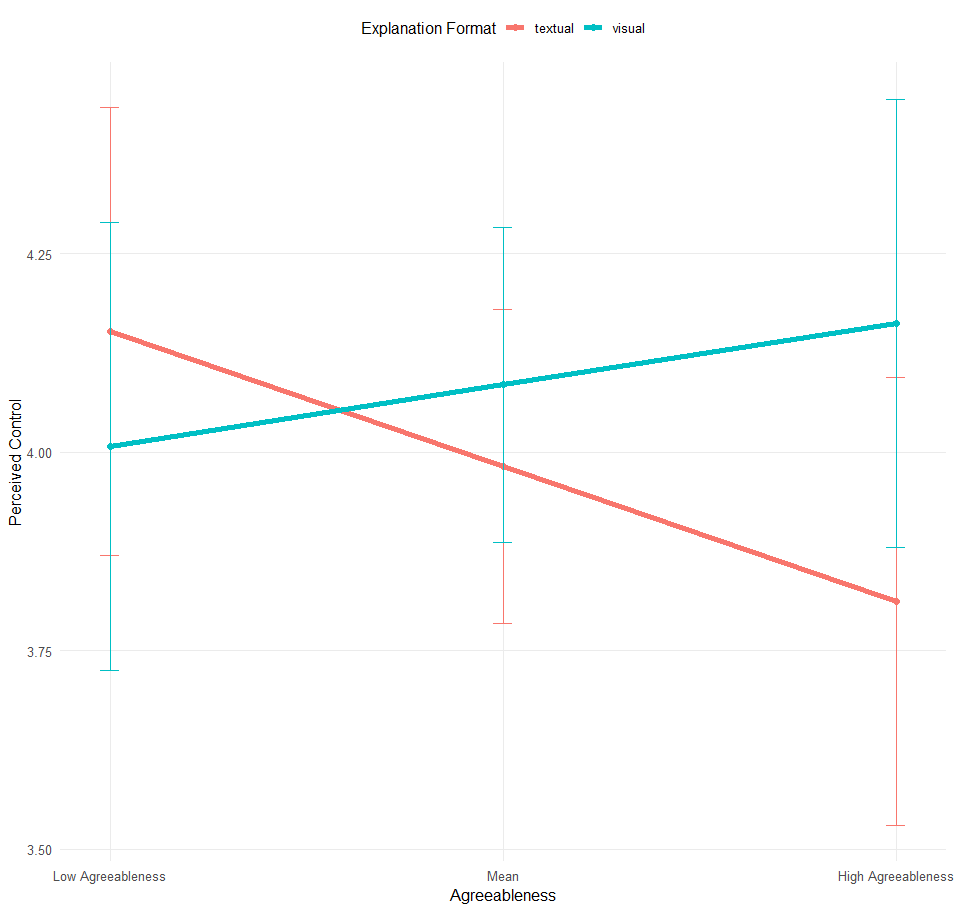}
    \Description{interaction plot}
    \caption{Agreeableness and Perceived Control}
    \label{fig:rq2_agree_control}
  \end{subfigure}\hfill
  \begin{subfigure}[t]{0.499\columnwidth}
    \centering
    \includegraphics[width=\linewidth]{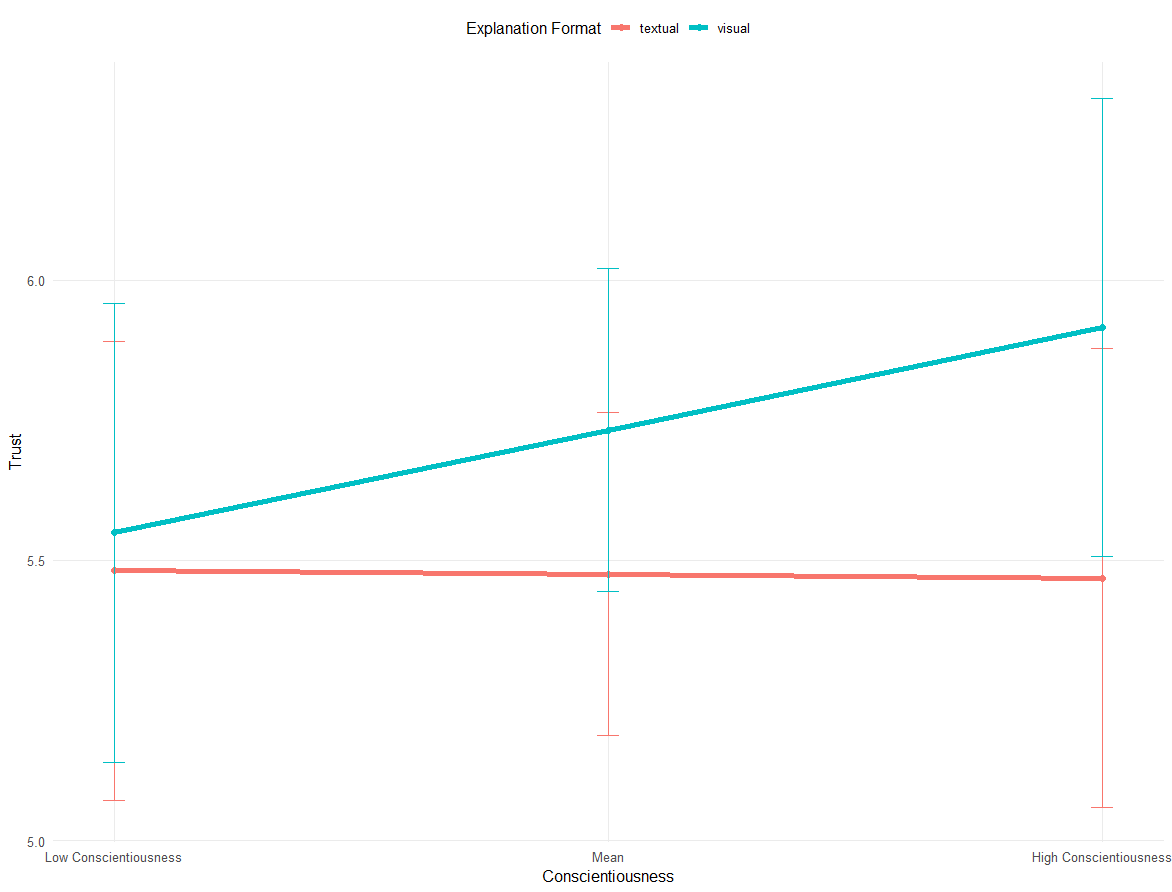}
    \caption{Conscientiousness and Trust}
    \label{fig:rq2-const_trust}
  \end{subfigure}
  \begin{subfigure}[t]{0.499\columnwidth}
    \centering
    \includegraphics[width=\linewidth]{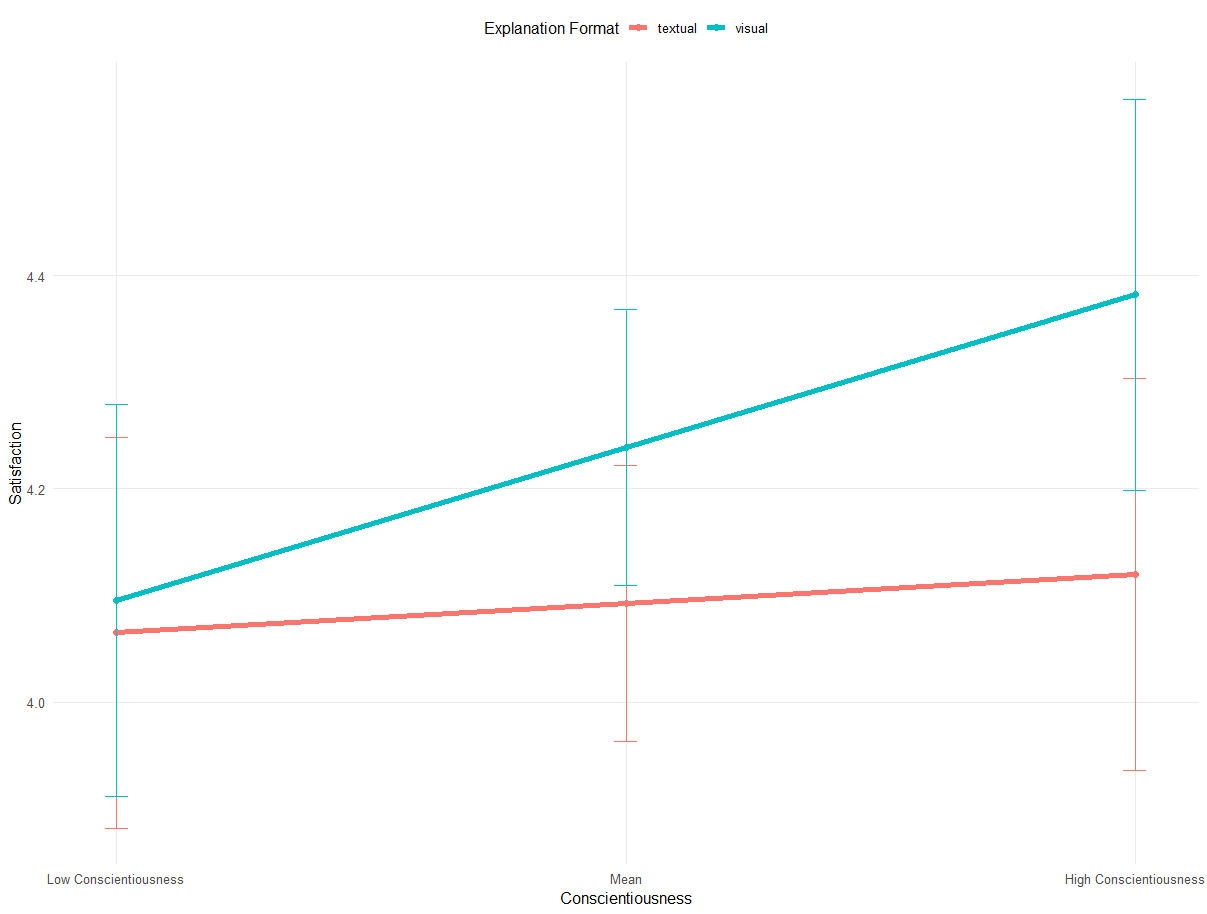}
    \caption{Conscientiousness and Satisfaction}
    \label{fig:rq2-const_sat}
  \end{subfigure}\hfill
  \begin{subfigure}[t]{0.499\columnwidth}
    \centering
    \includegraphics[width=\linewidth]{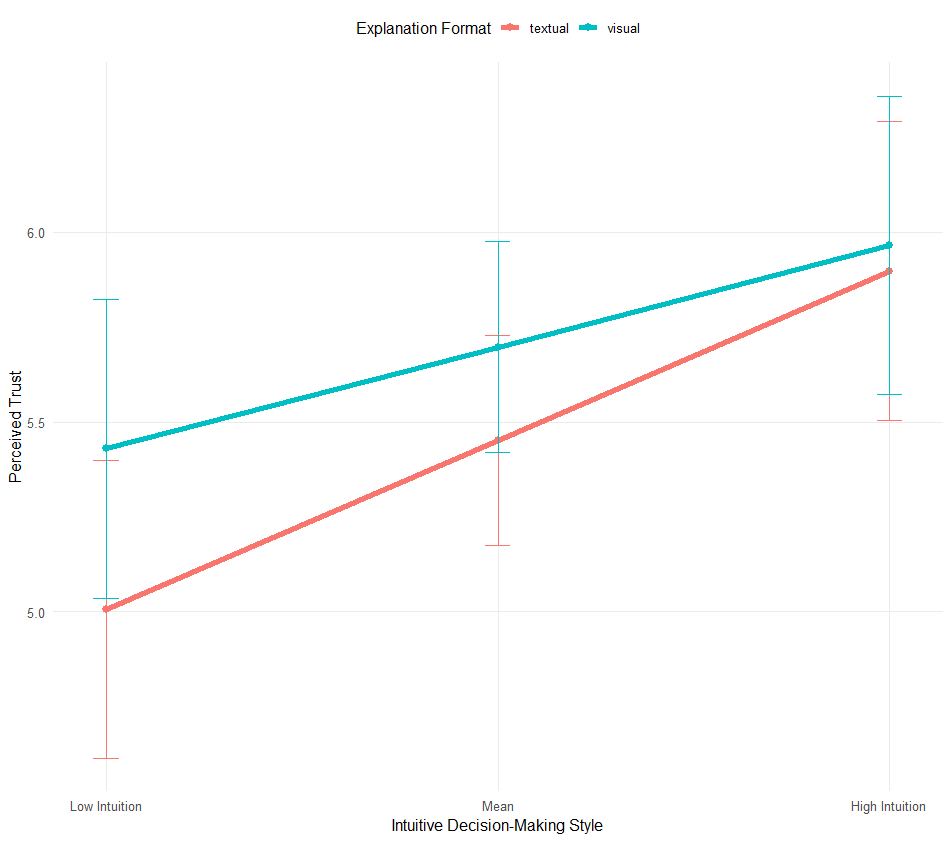}
    \caption{Intuitive DMS and Trust}
    \label{fig:rq2-DMS}
  \end{subfigure}
  \caption{The significant interaction effects between PCs and explanation formats}
  \label{fig:rq1-4subfigs}
\end{figure}

\textit{Big Five Traits:}
Agreeableness moderated the effect of explanation format on perceived control (b = 0.24, t = 2.07), with visual explanations increasing control for users high in Agreeableness (b = 0.35, p = .037), but not for users low or average in Agreeableness (Fig. \ref{fig:rq2_agree_control}). Conscientiousness moderated the effects of explanation format on trust (b = 0.17, t = 2.20) and satisfaction (b = 0.10, t = 2.18), such that visual explanations led to higher trust and satisfaction for users with average to high Conscientiousness, but not for users low in Conscientiousness (Figs. \ref{fig:rq2-const_trust}, \ref{fig:rq2-const_sat}). No moderation effects were observed for Extraversion, Emotional Stability, or Openness.

\textit{Decision Making Style (DMS):} Rational DMS did not moderate the effects of explanation format on any dependent variable. In contrast, intuitive DMS showed significant main effects on perceived transparency (b = 0.33, t = 3.31), trust (b = 0.58, t = 3.12), perceived control (b = 0.33, t = 2.44), and satisfaction (b = 0.22, t = 2.52). A significant interaction between explanation format and intuitive DMS was observed for trust ($b = −0.23, t = −2.23$), indicating that although visual explanations increased trust overall (b = 0.32, t = 2.85), this advantage decreased with higher intuitive DMS. Simple-effects analyses showed that visual explanations led to higher trust for users low ($\Delta = 0.42, z = −3.77, p < .001$) and average ($\Delta = 0.25, z = −3.11, p = .002$) in intuitive DMS, but not for highly intuitive users ($\Delta = 0.07, z = −0.61, p = .54$) (see Figure \ref{fig:rq2-DMS}).

\textit{Other PCs:} No moderation effects were observed for Need for Cognition, Visualization Familiarity, or Technical Expertise, as the interaction between explanation format and these characteristics was non-significant across all dependent variables. 
\subsection{Qualitative Analysis}
\subsubsection{Visual vs. Textual}
Answering the question, "Which explanation (textual or visual)
do you prefer, and why?", the majority of the users (n=46,) preferred visual explanations over textual ones because \textit{it is easy to understand} (n=14) and \textit{"it's very quick to understand"} (n=14).  
Furthermore, participants found visual explanations quick and faster (n=14) as they "\textit{do not require a lot of time and effort to process it}" (P34), conveyed information “\textit{at first glance} (P36, P51)", 
and \textit{"does not require clicking to get more information"} (P6, P36, P33).
The feature of visual explanations admired by most of the participants (n=17) is the use of consistent color-coding across all the visualizations, as they found it to be \textit{"attractive"} (P44, P43, P17), \textit{"pleasing to the eyes"} (P44, P45, P25), and "enjoyable" (P32).
Only few users (n=6) preferred textual explanations over visual, mentioning that "\textit{text is more clear, structured and simple}" (P19, P41, P46 P50), and "\textit{written information is more clear}" (P46).
Moreover, two participants desired to have a combination of both formats, for instance, "\textit{Textual for me is too much text, if we put some colors and visual elements in textual ones, it will be more helpful}" (P26).
\subsubsection{Perceived Control}
Regarding the question “Which explanation (textual or visual) gave you a sense of control with the ERS, and why?”, most participants (n=31) reported that visual explanations provided a stronger sense of control. They described visual explanations as more interactive, 
for instance, P52 explained, “\textit{you see a colored word and want to interact with it… when you hover over a word, it shows graphs and changing percentages, and the moving bars are very engaging}”. Similarly, P27 noted that, "\textit{if I want to know something about the colored words which are being shown to me, then I would interact with them}".
In contrast, some participants (n=11) felt that textual explanations fostered a greater sense of control as they required explicit interaction to reveal additional information, as P53 noted, “\textit{I feel more authority with playing with the information, and I can control the details I want to view or not}”. 
This perception was influenced by the textual explanation design, which relied on bold keywords rather than color-coding and required users to hover or click to access additional information.
Some participants (n=12) reported a similar sense of control for both formats, noting that they differed mainly in their use of color As P48 summarized, “\textit{I can hover and click in the same way and I understand the things}”. 
\subsubsection{Transparency}
In response to the question “Which explanation gave you a sense of transparency, and why?”, most participants (n=32) preferred visual explanations because they were easier and quicker to understand, which helped them better and faster grasp how recommendations were generated. For example, P30 mentioned, "\textit{it was more clear for me why the recommender chose this video}". Similarly, P43 stated; "\textit{it showed me everything clearly, to which topic was it more similar to ... It was more appealing, easy to interpret because of the graphs and the colors}". 
On the other hand, some participants (n=11) reported that textual explanations provided a stronger sense of transparency, highlighting that the greater amount of text conveyed a sense of richer detail. For example, “\textit{Because it’s more detailed, it shows everything and doesn’t feel like it’s hiding anything}” (P41). Furthermore, another group (n=11) perceived both explanation formats as equally transparent as “\textit{Both explanation formats reveal the same information}” (P51). 
With respect to explanation intelligibility types, participants
consistently valued "what" and "why" (abstract) explanations, as these
helped them understand which concepts influenced the recommendations. Visual cues such as color band with similarity percentages and color-coded keyphrases were particularly appreciated. In
contrast, most participants (n=36) reported that why (detailed)
explanations were unnecessary, describing them as "too much information" and they simply "do not need to know this detail" while making a decision to watch a video. 
\subsubsection{Trust}
In response to the question “Which explanation gave you a sense of trust with the system, and why?”, participants expressed mixed opinions. Most participants (n=23) reported that visual explanations fostered a stronger sense of trust in the system as they were more transparent and easier to understand, which made them feel more comfortable and confident in the system. For example, P38 expressed: \textit{"It's more transparent, so I trust it more"} 
Similarly, P48 explained, \textit{“I now know which concepts the system uses and how similar a video is to my concepts. I feel like I know everything, and that makes me trust the system and its recommendations”}. Additionally, P45 highlighted, "because I understand it more, and I can feel comfortable with it". Other participants found visual explanations to be helpful in faster decision making which increased their trust in the system. For example, P9 pointed out: \textit{clicking on the keyphrases, and being able to know how close they can be to the concept, that will help me to directly select which video to watch}".
Another substantial group of participants (n=18) reported that both explanation formats fostered a similar sense of trust as it was driven by the information conveyed, the quality of the recommendations, or the support in decision making rather than by how easily or quickly the explanation could be understood. As 
P50 explained, \textit{“Visual was just quicker, but when it comes to trust, it’s not about how quick it is to digest, but how accurate the results are”}. 
A subset of participants (n=11) reported higher perceived trust with the ERS having explanations in textual format.  
These participants emphasized that textual explanations provided more complete and explicit information, which reduced ambiguity and the need for interpretation.
For instance, 
P43 explained, "\textit{it provides more detail so it gains my trust as I feel the system is telling me everything clearly. In visual you can be a little confused in interpreting the visual clues, but in textual everything is written and you don't have to guess the meaning of a visualization}”. 
Others directly associated words with credibility, with P52 stating that, “\textit{the written information gives you a feeling that words are more trustable than visual}”. 
\subsubsection{Satisfaction}
In response to the question “Which explanation gave you a sense of satisfaction with the system, and why?”, the majority of participants (n=45) preferred visual explanations. For many participants, satisfaction was driven by how effectively the explanation helped them identify a relevant video, with visual explanations supporting efficient and more confident selection.
For instance, P37 noted that visuals helped them to \textit{“make decisions faster, resulting in a better experience"}. Similarly, P8 explained, “\textit{I can use the color band to find the desired video quickly}”, 
and P26 summarized as visual explanation helping them \textit{“find a video that will help me understand a concept”}.
Another reason mentioned by many participants was visual explanations being quicker to understand. 
For instance, P5 described visual explanations as \textit{“quick to understand”}, while P4 highlighted how visual elements such as, \textit{"the word cloud and color bands made key information visible right away"}. This sentiment was echoed by P36, who preferred visuals because they wanted to \textit{“directly play the video without first reading extensive text"}. 
Another reason was that visual explanations improved satisfaction by making the system’s reasoning more transparent 
as P54 noted that, \textit{"understanding why the recommendation was made helped them feel satisfied with it"}.
Finally, while some found them \textit{“comfortable to use"} (P27, P30), others described them as \textit{“more interactive and colorful”} (P38), and finding \textit{"the graphs and word clouds useful"} (P42). 
Related to satisfaction, textual explanations were preferred by a small subset of participants (n=8) as they were perceived as clearer, more detailed, and more reliable for decision making. While P19 described them as \textit{“more clear and professional”}, P41 found them \textit{“simple, and less colored”}. Some participants expressed concerns that visual explanations could be confusing despite being engaging, as P43 explained, “\textit{visual is catchy ... but you can be confused why it is so”}. Some participants were satisfied with the amount of detail provided by the textual explanation. For example, P46 emphasized that, \textit{“textual is explaining in a lot more detail”}. Other participants related satisfaction to the ability to make good decisions. For instance, P7 stated that the textual explanation \textit{“helped me to find useful videos”}.
\section{Discussion}
In this section, we discuss the main findings of our study in relation to our research questions and provide some guidelines for the effective design of explanations in ERSs.
\subsection{Visual vs. Textual}
Our qualitative findings indicate a strong preference for visual explanations, as simple visual cues such as color-coding and interactive features (e.g., clicks and hover effects) enabled users to quickly and effortlessly understand the reasoning behind recommendations. Participants particularly valued that key explanatory information was presented upfront and could be grasped at a glance, supporting rapid decision making. This preference aligns with the inherent advantages of visual representations, which are generally processed more efficiently than textual information and support faster comprehension with lower cognitive effort \cite{munzner2025visualization, ware2019information}. 
Our findings are consistent with earlier work by \citet{kouki2017user}, who reported that users tend to prefer simple visual formats over more complex ones.
This also aligns with prior work pointing out that end-users do not necessarily benefit from highly exploratory, information-heavy, or overly complex visual interfaces, even when such interfaces are visually sophisticated \cite{ooge2022explaining}. 
\textbf{Design guideline}: In line with established principles of data visualization, simplicity plays a critical role in the design of visual explanations as well. It is essential to provide visual explanations as simple as possible, yet with
enough intuitive interaction mechanisms to allow users to quickly build an accurate mental model of how the ERS works.
\subsection{Perceived Control}
Perceived control in RSs refers to users’ feeling that they can influence the system’s behavior and actively engage with the information provided \cite{pu2012evaluating, Tintarev2015}. Our findings indicate that visual explanations often fostered a stronger sense of control due to their interaction features, such as colors, clicks, hover effects, and dynamic visual elements, which encouraged exploration and active engagement. Participants felt that these interactions allowed them to control the explanation and decide what information to view and when. These findings are in line with earlier research on explainable recommendation suggesting that perceived control increases when users can actively decide when and how much explanatory information to access, which reduces cognitive effort, rather than being presented with all information upfront \cite{guesmi2021demand, guesmi2024interactive,millecamp2019explain,Tintarev2015}.
This echoes the suggestion in \cite{chatti2024visualization} to provide layered visual explanations that follow a "Basic Explanation – Show the Important – Details on Demand" approach to help users iteratively build better mental models of how the RS works. This further aligns with suggestions in broader XAI research stressing that the selective characteristic of explanation
needs to be taken into account in order to achieve meaningful explanation \cite{miller2019explanation}. 
\textbf{Design guideline:} Provide interactive visual explanations in ERSs that support selective and on-demand access to explanatory information, based on user's needs. 
\subsection{Transparency}
Transparency in RS is related to the capability of a system to expose the reasoning
behind a recommendation to its users \cite{herlocker2000explaining} and is defined as users’ understanding of the RS’s inner logic \cite{tintarev2007survey,pu2012evaluating}. 
Our study shows that visual explanations provided better sense of transparency with the ERS, mainly because they were  easier and faster to understand that how recommendations were generated. Moreover, users found the "what" and "why (abstract)" explanations enough to understand (1) what data does the system use and (2) why and how well does a recommended item fit one’s preferences. By contrast, they found the "why (detailed)" explanation unnecessary, as it contains too much information that is not needed to make a decision. Our findings indicate that, in an ERS, an explanation does not need to be sound (i.e., the extent to which the explanation is truthful in describing the underlying system \cite{kulesza2015principles}) or complete (i.e., the extent to which all of the underlying system is described by the explanation \cite{kulesza2015principles}). What is more important is that the explanation remains comprehensible to avoid overwhelming users.
This is consistent with results of previous research on XAI showing that for specific user groups, detailed explanation is often not needed because the provision of additional explanations increases cognitive effort \cite{kizilcec2016much, kulesza2013too, yang2020visual}.
This is also in line with the suggestion provided by \citet{kizilcec2016much} who concluded that designing for effectiveness requires balanced interface transparency, i.e., “not too little and not too much”.
\textbf{Design guideline:} For increased transparency in ERSs, it is important to provide “what” and “why” visual explanations that clearly and intuitively communicate how user preferences are linked to recommendations, through visually interpretable representations that support users’ understanding of the system’s reasoning. Moreover, it is essential to provide a visual explanation with just the right amount of information which is "not too little and not too much" to allow users to build accurate mental models of how the ERS works, without overwhelming them.
\subsection{Trust}
Trust has long been recognized as a key factor in explainable recommendation. \citet{tintarev2007survey} conceptualize trust as "increasing users’ confidence in the system", while \citet{pu2011user} frame it as part of users’ overall attitude toward the system. More broadly, trust in RS can be defined as the extent to which users are confident in
and willing to act on the basis of the recommendations (adapted from \cite{yang2020visual,madsen2000measuring}).
Based on the alignment between the perceived and actual performance of the system, \citet{yang2020visual} distinguished between appropriate trust (i.e., to [not] follow an [in]correct recommendation), overtrust, and undertrust. 
Our study indicates that both visual and textual explanations fostered users’ appropriate trust (i.e., users’ ability to rely on the ERS when it is correct and to recognize when it is incorrect) and facilitated the decision making process (i.e., whether to watch a recommended video or not). However, the visual explanation yielded significantly higher appropriate trust and faster decisions than the textual one.
Our results further show that both explanation formats increased users’ trust in the ERS, because they are transparent and easy to understand. This confirms prior findings that providing transparency could enhance
users’ trust in the RS \cite{hellmann2022development,  Tintarev2015, nunes2017systematic, pu2012evaluating,kunkel2019let,ooge2022explaining,guesmijustification2023} and that transparency and trust are often linked, following the intuition that users are more likely to trust systems they can understand than one that is a black box \cite{siepmann2023trust}.
Further, we found that trust in an ERS can also depend on the system’s ability to provide accurate and useful recommendations. This outcome is consistent with prior research highlighting that users' trust in the RS might be influenced by its ability to formulate good recommendations \cite{pu2011user} and the accuracy
of the recommendation algorithm \cite{tintarev2010designing}.
Overall, these different perspectives confirm that trust is a multi-faceted concept influenced by multiple factors, as also highlighted in prior research on explainable recommendation \cite{siepmann2023trust} and XAI  \cite{yang2020visual,miller2022we, liao2022designing}. 
\textbf{Design guideline:} Provide visual explanations that are easy to understand to foster appropriate trust and efficient decision making in ERSs. 
\subsection{Satisfaction}
Satisfaction (i.e., increase the
ease of use or enjoyment \cite{Tintarev2015}) determines what users think and feel while using an RS \cite{pu2011user}. Our results show a wide agreement in favor of visual explanations fostering a better sense of satisfaction with the ERS.   
The visual explanation was perceived as more engaging and comfortable to interact with, easier to understand, and more effective to make a faster and confident decision on the usefulness of a recommendation, which contributed to increased satisfaction with the ERS. This indicates a positive association between ease to use, usefulness, and satisfaction. This aligns with the view on satisfaction presented in \cite{nunes2017systematic}, noting that satisfaction is not considered as a single goal, but can be split into sub-goals of ease to use and usefulness.  
Moreover, our results show a positive correlation
between transparency and satisfaction. 
This is in line with earlier studies which found that the user’s overall satisfaction with an RS is assumed to be strongly related to transparency \cite{Tintarev2015, guesmijustification2023, gedikli2014should, balog2020measuring, guesmi2024interactive}.
\textbf{Design guideline:} To foster user satisfaction with the ERS, visual explanations should be easy to use and understand, engaging, comfortable to interact with, and supportive in quick decision making.
\subsection{Personal Characteristics}
Our study shows that visual explanations significantly increased perceived control for users higher in Agreeableness, as well as trust and satisfaction for users higher in Conscientiousness. Moreover, we found that 
users with high intuitive DMS benefited equally from both formats. One possible reason that might result in these preferences is that agreeable and conscientious users benefit from the clarity and structure provided by visual explanations, whereas, users with intuitive DMS who rely 
on first impressions when making decisions
found both explanation formats intuitive enough to support immediate observability.
However, despite the significant effects that emerged in relation to Agreeableness, Conscientiousness, and intuitive DMS, the overall pattern remained consistent with the main effects observed across all users, namely that, compared to textual explanations, visual explanations led to higher levels of trust and satisfaction (both with significance), perceived control, and transparency. This suggests that personalizing explanation formats based on Big Five traits may not be necessary. 
Furthermore, our results show no moderating effects of other PCs (i.e., NFC, VF, and TE) on users’ perceptions of explanation formats. A possible explanation for the absence of these moderation effects lies in the deliberately simple and intuitive design of the explanations. Both explanations presented information clearly, did not require any complicated interaction, and avoided technical complexity. As a result, understanding the explanations did not demand additional cognitive effort, familiarity with advanced visualizations, or prior knowledge of RS.
\textbf{Design guideline:} Visual explanations in ERSs that follow a clear, intuitive, and simple design would likely be sufficient for users with diverse PCs to foster perceived control, transparency, trust, and satisfaction. 
\section{Conclusion}
In this study, we investigated how explanation format (visual vs. textual) and users’ personal characteristics (PCs) impact their perceptions of control, transparency, trust, and satisfaction in an educational recommender system (ERS), an area that still remained underexplored. Through a within-subject user study (n=54), we found that visual explanations significantly increase trust and satisfaction, while perceived control and transparency remain comparable across explanation formats.
Furthermore, PCs played a limited role in shaping users’ perceptions of explanation formats, with visual explanations yielding consistent benefits across users.
Based on our findings, we derived design guidelines for the effective design of explanations in ERSs. 


\begin{thebibliography}{74}


\ifx \showCODEN    \undefined \def \showCODEN     #1{\unskip}     \fi
\ifx \showISBNx    \undefined \def \showISBNx     #1{\unskip}     \fi
\ifx \showISBNxiii \undefined \def \showISBNxiii  #1{\unskip}     \fi
\ifx \showISSN     \undefined \def \showISSN      #1{\unskip}     \fi
\ifx \showLCCN     \undefined \def \showLCCN      #1{\unskip}     \fi
\ifx \shownote     \undefined \def \shownote      #1{#1}          \fi
\ifx \showarticletitle \undefined \def \showarticletitle #1{#1}   \fi
\ifx \showURL      \undefined \def \showURL       {\relax}        \fi
\providecommand\bibfield[2]{#2}
\providecommand\bibinfo[2]{#2}
\providecommand\natexlab[1]{#1}
\providecommand\showeprint[2][]{arXiv:#2}

\bibitem[Adadi and Berrada(2018)]%
        {adadi2018peeking}
\bibfield{author}{\bibinfo{person}{Amina Adadi} {and} \bibinfo{person}{Mohammed Berrada}.} \bibinfo{year}{2018}\natexlab{}.
\newblock \showarticletitle{Peeking inside the black-box: a survey on explainable artificial intelligence (XAI)}.
\newblock \bibinfo{journal}{\emph{IEEE access}}  \bibinfo{volume}{6} (\bibinfo{year}{2018}), \bibinfo{pages}{52138--52160}.
\newblock


\bibitem[Ain et~al\mbox{.}(2022)]%
        {ain2022multi}
\bibfield{author}{\bibinfo{person}{Qurat~Ul Ain}, \bibinfo{person}{Mohamed~Amine Chatti}, \bibinfo{person}{Mouadh Guesmi}, {and} \bibinfo{person}{Shoeb Joarder}.} \bibinfo{year}{2022}\natexlab{}.
\newblock \showarticletitle{A multi-dimensional conceptualization framework for personalized explanations in recommender systems}. In \bibinfo{booktitle}{\emph{Companion Proceedings of the 27th International Conference on Intelligent User Interfaces}}. \bibinfo{pages}{11--23}.
\newblock


\bibitem[Al-Hazwani et~al\mbox{.}(2025)]%
        {al2025f}
\bibfield{author}{\bibinfo{person}{Ibrahim Al-Hazwani}, \bibinfo{person}{Gabriela Morgenshtern}, \bibinfo{person}{Mennatallah El-Assady}, {and} \bibinfo{person}{J{\"u}rgen Bernard}.} \bibinfo{year}{2025}\natexlab{}.
\newblock \showarticletitle{f-RecX: A framework for designing effective textual explanations in recommender systems’ user interfaces}.
\newblock \bibinfo{journal}{\emph{International Journal of Human-Computer Studies}} (\bibinfo{year}{2025}), \bibinfo{pages}{103627}.
\newblock


\bibitem[Balog and Radlinski(2020)]%
        {balog2020measuring}
\bibfield{author}{\bibinfo{person}{Krisztian Balog} {and} \bibinfo{person}{Filip Radlinski}.} \bibinfo{year}{2020}\natexlab{}.
\newblock \showarticletitle{Measuring recommendation explanation quality: The conflicting goals of explanations}. In \bibinfo{booktitle}{\emph{Proceedings of the 43rd international ACM SIGIR conference on research and development in information retrieval}}. \bibinfo{pages}{329--338}.
\newblock


\bibitem[Braun and Clarke(2021)]%
        {braun2021thematic}
\bibfield{author}{\bibinfo{person}{Virginia Braun} {and} \bibinfo{person}{Victoria Clarke}.} \bibinfo{year}{2021}\natexlab{}.
\newblock \showarticletitle{Thematic analysis: A practical guide}.
\newblock  (\bibinfo{year}{2021}).
\newblock


\bibitem[Chang et~al\mbox{.}(2016)]%
        {chang2016crowd}
\bibfield{author}{\bibinfo{person}{Shuo Chang}, \bibinfo{person}{F~Maxwell Harper}, {and} \bibinfo{person}{Loren~Gilbert Terveen}.} \bibinfo{year}{2016}\natexlab{}.
\newblock \showarticletitle{Crowd-based personalized natural language explanations for recommendations}. In \bibinfo{booktitle}{\emph{Proceedings of the 10th ACM conference on recommender systems}}. \bibinfo{pages}{175--182}.
\newblock


\bibitem[Chatti et~al\mbox{.}(2024)]%
        {chatti2024visualization}
\bibfield{author}{\bibinfo{person}{Mohamed~Amine Chatti}, \bibinfo{person}{Mouadh Guesmi}, {and} \bibinfo{person}{Arham Muslim}.} \bibinfo{year}{2024}\natexlab{}.
\newblock \showarticletitle{Visualization for recommendation explainability: a survey and new perspectives}.
\newblock \bibinfo{journal}{\emph{ACM Transactions on Interactive Intelligent Systems}} \bibinfo{volume}{14}, \bibinfo{number}{3} (\bibinfo{year}{2024}), \bibinfo{pages}{1--40}.
\newblock


\bibitem[Chatti et~al\mbox{.}(2022)]%
        {chatti2022more}
\bibfield{author}{\bibinfo{person}{Mohamed~Amine Chatti}, \bibinfo{person}{Mouadh Guesmi}, \bibinfo{person}{Laura Vorgerd}, \bibinfo{person}{Thao Ngo}, \bibinfo{person}{Shoeb Joarder}, \bibinfo{person}{Qurat~Ul Ain}, {and} \bibinfo{person}{Arham Muslim}.} \bibinfo{year}{2022}\natexlab{}.
\newblock \showarticletitle{Is more always better? The effects of personal characteristics and level of detail on the perception of explanations in a recommender system}. In \bibinfo{booktitle}{\emph{Proceedings of the 30th ACM Conference on User Modeling, Adaptation and Personalization}}. \bibinfo{pages}{254--264}.
\newblock


\bibitem[Chen et~al\mbox{.}(2019)]%
        {chen2019personalized}
\bibfield{author}{\bibinfo{person}{Xu Chen}, \bibinfo{person}{Hanxiong Chen}, \bibinfo{person}{Hongteng Xu}, \bibinfo{person}{Yongfeng Zhang}, \bibinfo{person}{Yixin Cao}, \bibinfo{person}{Zheng Qin}, {and} \bibinfo{person}{Hongyuan Zha}.} \bibinfo{year}{2019}\natexlab{}.
\newblock \showarticletitle{Personalized fashion recommendation with visual explanations based on multimodal attention network: Towards visually explainable recommendation}. In \bibinfo{booktitle}{\emph{Proceedings of the 42nd international ACM SIGIR conference on research and development in information retrieval}}. \bibinfo{pages}{765--774}.
\newblock


\bibitem[Costa et~al\mbox{.}(2018)]%
        {costa2018automatic}
\bibfield{author}{\bibinfo{person}{Felipe Costa}, \bibinfo{person}{Sixun Ouyang}, \bibinfo{person}{Peter Dolog}, {and} \bibinfo{person}{Aonghus Lawlor}.} \bibinfo{year}{2018}\natexlab{}.
\newblock \showarticletitle{Automatic generation of natural language explanations}. In \bibinfo{booktitle}{\emph{Companion Proceedings of the 23rd International Conference on Intelligent User Interfaces}}. \bibinfo{pages}{1--2}.
\newblock


\bibitem[Dominguez et~al\mbox{.}(2019)]%
        {dominguez2019effect}
\bibfield{author}{\bibinfo{person}{Vicente Dominguez}, \bibinfo{person}{Pablo Messina}, \bibinfo{person}{Ivania Donoso-Guzm{\'a}n}, {and} \bibinfo{person}{Denis Parra}.} \bibinfo{year}{2019}\natexlab{}.
\newblock \showarticletitle{The effect of explanations and algorithmic accuracy on visual recommender systems of artistic images}. In \bibinfo{booktitle}{\emph{Proceedings of the 24th International Conference on Intelligent User Interfaces}}. \bibinfo{pages}{408--416}.
\newblock


\bibitem[Faul et~al\mbox{.}(2007)]%
        {faul2007g}
\bibfield{author}{\bibinfo{person}{Franz Faul}, \bibinfo{person}{Edgar Erdfelder}, \bibinfo{person}{Albert-Georg Lang}, {and} \bibinfo{person}{Axel Buchner}.} \bibinfo{year}{2007}\natexlab{}.
\newblock \showarticletitle{G* Power 3: A flexible statistical power analysis program for the social, behavioral, and biomedical sciences}.
\newblock \bibinfo{journal}{\emph{Behavior research methods}} \bibinfo{volume}{39}, \bibinfo{number}{2} (\bibinfo{year}{2007}), \bibinfo{pages}{175--191}.
\newblock


\bibitem[Gedikli et~al\mbox{.}(2014)]%
        {gedikli2014should}
\bibfield{author}{\bibinfo{person}{Fatih Gedikli}, \bibinfo{person}{Dietmar Jannach}, {and} \bibinfo{person}{Mouzhi Ge}.} \bibinfo{year}{2014}\natexlab{}.
\newblock \showarticletitle{How should I explain? A comparison of different explanation types for recommender systems}.
\newblock \bibinfo{journal}{\emph{International Journal of Human-Computer Studies}} \bibinfo{volume}{72}, \bibinfo{number}{4} (\bibinfo{year}{2014}), \bibinfo{pages}{367--382}.
\newblock


\bibitem[Gosling et~al\mbox{.}(2003)]%
        {gosling2003very}
\bibfield{author}{\bibinfo{person}{Samuel~D Gosling}, \bibinfo{person}{Peter~J Rentfrow}, {and} \bibinfo{person}{William~B Swann~Jr}.} \bibinfo{year}{2003}\natexlab{}.
\newblock \showarticletitle{A very brief measure of the Big-Five personality domains}.
\newblock \bibinfo{journal}{\emph{Journal of Research in personality}} \bibinfo{volume}{37}, \bibinfo{number}{6} (\bibinfo{year}{2003}), \bibinfo{pages}{504--528}.
\newblock


\bibitem[Guesmi et~al\mbox{.}(2024)]%
        {guesmi2024interactive}
\bibfield{author}{\bibinfo{person}{Mouadh Guesmi}, \bibinfo{person}{Mohamed~Amine Chatti}, \bibinfo{person}{Shoeb Joarder}, \bibinfo{person}{Qurat~Ul Ain}, \bibinfo{person}{Rawaa Alatrash}, \bibinfo{person}{Clara Siepmann}, {and} \bibinfo{person}{Tannaz Vahidi}.} \bibinfo{year}{2024}\natexlab{}.
\newblock \showarticletitle{Interactive explanation with varying level of details in an explainable scientific literature recommender system}.
\newblock \bibinfo{journal}{\emph{International Journal of Human--Computer Interaction}} \bibinfo{volume}{40}, \bibinfo{number}{22} (\bibinfo{year}{2024}), \bibinfo{pages}{7248--7269}.
\newblock


\bibitem[Guesmi et~al\mbox{.}(2023)]%
        {guesmijustification2023}
\bibfield{author}{\bibinfo{person}{Mouadh Guesmi}, \bibinfo{person}{Mohamed~Amine Chatti}, \bibinfo{person}{Shoeb Joarder}, \bibinfo{person}{Qurat~Ul Ain}, \bibinfo{person}{Clara Siepmann}, \bibinfo{person}{Hoda Ghanbarzadeh}, {and} \bibinfo{person}{Rawaa Alatrash}.} \bibinfo{year}{2023}\natexlab{}.
\newblock \showarticletitle{Justification vs. Transparency: Why and How Visual Explanations in a Scientific Literature Recommender System}.
\newblock \bibinfo{journal}{\emph{Information}} \bibinfo{volume}{14}, \bibinfo{number}{7} (\bibinfo{year}{2023}).
\newblock
\showISSN{2078-2489}
\href{https://doi.org/10.3390/info14070401}{doi:\nolinkurl{10.3390/info14070401}}


\bibitem[Guesmi et~al\mbox{.}(2021b)]%
        {guesmi2021demand}
\bibfield{author}{\bibinfo{person}{Mouadh Guesmi}, \bibinfo{person}{Mohamed~Amine Chatti}, \bibinfo{person}{Laura Vorgerd}, \bibinfo{person}{Shoeb Joarder}, \bibinfo{person}{Shadi Zumor}, \bibinfo{person}{Yiqi Sun}, \bibinfo{person}{Fangzheng Ji}, {and} \bibinfo{person}{Arham Muslim}.} \bibinfo{year}{2021}\natexlab{b}.
\newblock \showarticletitle{On-demand personalized explanation for transparent recommendation}. In \bibinfo{booktitle}{\emph{Adjunct Proceedings of the 29th ACM Conference on User Modeling, Adaptation and Personalization}}. \bibinfo{pages}{246--252}.
\newblock


\bibitem[Guesmi et~al\mbox{.}(2021a)]%
        {guesmi2021input}
\bibfield{author}{\bibinfo{person}{Mouadh Guesmi}, \bibinfo{person}{Mohamed~Amine Chatti}, \bibinfo{person}{Laura Vorgerd}, \bibinfo{person}{Shoeb~Ahmed Joarder}, \bibinfo{person}{Qurat~Ul Ain}, \bibinfo{person}{Thao Ngo}, \bibinfo{person}{Shadi Zumor}, \bibinfo{person}{Yiqi Sun}, \bibinfo{person}{Fangzheng Ji}, {and} \bibinfo{person}{Arham Muslim}.} \bibinfo{year}{2021}\natexlab{a}.
\newblock \showarticletitle{Input or Output: Effects of Explanation Focus on the Perception of Explainable Recommendation with Varying Level of Details.}. In \bibinfo{booktitle}{\emph{Intrs@ recsys}}. \bibinfo{pages}{55--72}.
\newblock


\bibitem[Guesmi et~al\mbox{.}(2022)]%
        {guesmi2022explaining}
\bibfield{author}{\bibinfo{person}{Mouadh Guesmi}, \bibinfo{person}{Mohamed~Amine Chatti}, \bibinfo{person}{Laura Vorgerd}, \bibinfo{person}{Thao Ngo}, \bibinfo{person}{Shoeb Joarder}, \bibinfo{person}{Qurat~Ul Ain}, {and} \bibinfo{person}{Arham Muslim}.} \bibinfo{year}{2022}\natexlab{}.
\newblock \showarticletitle{Explaining user models with different levels of detail for transparent recommendation: A user study}. In \bibinfo{booktitle}{\emph{Adjunct Proceedings of the 30th ACM Conference on User Modeling, Adaptation and Personalization}}. \bibinfo{pages}{175--183}.
\newblock


\bibitem[Hada et~al\mbox{.}(2021)]%
        {hada2021rexplug}
\bibfield{author}{\bibinfo{person}{Deepesh~V Hada}, \bibinfo{person}{Vijaikumar M}, {and} \bibinfo{person}{Shirish~K Shevade}.} \bibinfo{year}{2021}\natexlab{}.
\newblock \showarticletitle{Rexplug: Explainable recommendation using plug-and-play language model}. In \bibinfo{booktitle}{\emph{Proceedings of the 44th international ACM SIGIR conference on research and development in information retrieval}}. \bibinfo{pages}{81--91}.
\newblock


\bibitem[Hamilton et~al\mbox{.}(2016)]%
        {hamilton2016development}
\bibfield{author}{\bibinfo{person}{Katherine Hamilton}, \bibinfo{person}{Shin-I Shih}, {and} \bibinfo{person}{Susan Mohammed}.} \bibinfo{year}{2016}\natexlab{}.
\newblock \showarticletitle{The development and validation of the rational and intuitive decision styles scale}.
\newblock \bibinfo{journal}{\emph{Journal of personality assessment}} \bibinfo{volume}{98}, \bibinfo{number}{5} (\bibinfo{year}{2016}), \bibinfo{pages}{523--535}.
\newblock


\bibitem[Haugtvedt et~al\mbox{.}(1992)]%
        {haugtvedt1992need}
\bibfield{author}{\bibinfo{person}{Curtis~P Haugtvedt}, \bibinfo{person}{Richard~E Petty}, {and} \bibinfo{person}{John~T Cacioppo}.} \bibinfo{year}{1992}\natexlab{}.
\newblock \showarticletitle{Need for cognition and advertising: Understanding the role of personality variables in consumer behavior}.
\newblock \bibinfo{journal}{\emph{Journal of consumer psychology}} \bibinfo{volume}{1}, \bibinfo{number}{3} (\bibinfo{year}{1992}), \bibinfo{pages}{239--260}.
\newblock


\bibitem[Hellmann et~al\mbox{.}(2022)]%
        {hellmann2022development}
\bibfield{author}{\bibinfo{person}{Marco Hellmann}, \bibinfo{person}{Diana~C Hernandez-Bocanegra}, {and} \bibinfo{person}{J{\"u}rgen Ziegler}.} \bibinfo{year}{2022}\natexlab{}.
\newblock \showarticletitle{Development of an instrument for measuring users’ perception of transparency in recommender systems}.
\newblock \bibinfo{journal}{\emph{system}}  \bibinfo{volume}{12} (\bibinfo{year}{2022}), \bibinfo{pages}{7}.
\newblock


\bibitem[Herlocker et~al\mbox{.}(2000)]%
        {herlocker2000explaining}
\bibfield{author}{\bibinfo{person}{Jonathan~L Herlocker}, \bibinfo{person}{Joseph~A Konstan}, {and} \bibinfo{person}{John Riedl}.} \bibinfo{year}{2000}\natexlab{}.
\newblock \showarticletitle{Explaining collaborative filtering recommendations}. In \bibinfo{booktitle}{\emph{Proceedings of the 2000 ACM conference on Computer supported cooperative work}}. \bibinfo{pages}{241--250}.
\newblock


\bibitem[Hernandez-Bocanegra and Ziegler(2021)]%
        {hernandez2021explaining}
\bibfield{author}{\bibinfo{person}{Diana~C Hernandez-Bocanegra} {and} \bibinfo{person}{J{\"u}rgen Ziegler}.} \bibinfo{year}{2021}\natexlab{}.
\newblock \showarticletitle{Explaining review-based recommendations: Effects of profile transparency, presentation style and user characteristics}.
\newblock \bibinfo{journal}{\emph{i-com}} \bibinfo{volume}{19}, \bibinfo{number}{3} (\bibinfo{year}{2021}), \bibinfo{pages}{181--200}.
\newblock


\bibitem[Howell(1992)]%
        {howell1992statistical}
\bibfield{author}{\bibinfo{person}{David~C Howell}.} \bibinfo{year}{1992}\natexlab{}.
\newblock \bibinfo{booktitle}{\emph{Statistical methods for psychology}}.
\newblock \bibinfo{publisher}{PWS-Kent Publishing Co}.
\newblock


\bibitem[Kim et~al\mbox{.}(2018)]%
        {kim2018textual}
\bibfield{author}{\bibinfo{person}{Jinkyu Kim}, \bibinfo{person}{Anna Rohrbach}, \bibinfo{person}{Trevor Darrell}, \bibinfo{person}{John Canny}, {and} \bibinfo{person}{Zeynep Akata}.} \bibinfo{year}{2018}\natexlab{}.
\newblock \showarticletitle{Textual explanations for self-driving vehicles}. In \bibinfo{booktitle}{\emph{Proceedings of the European conference on computer vision (ECCV)}}. \bibinfo{pages}{563--578}.
\newblock


\bibitem[Kizilcec(2016)]%
        {kizilcec2016much}
\bibfield{author}{\bibinfo{person}{Ren{\'e}~F Kizilcec}.} \bibinfo{year}{2016}\natexlab{}.
\newblock \showarticletitle{How much information? Effects of transparency on trust in an algorithmic interface}. In \bibinfo{booktitle}{\emph{Proceedings of the 2016 CHI conference on human factors in computing systems}}. \bibinfo{pages}{2390--2395}.
\newblock


\bibitem[Knijnenburg et~al\mbox{.}(2012)]%
        {knijnenburg2012explaining}
\bibfield{author}{\bibinfo{person}{Bart~P Knijnenburg}, \bibinfo{person}{Martijn~C Willemsen}, \bibinfo{person}{Zeno Gantner}, \bibinfo{person}{Hakan Soncu}, {and} \bibinfo{person}{Chris Newell}.} \bibinfo{year}{2012}\natexlab{}.
\newblock \showarticletitle{Explaining the user experience of recommender systems}.
\newblock \bibinfo{journal}{\emph{User modeling and user-adapted interaction}} \bibinfo{volume}{22}, \bibinfo{number}{4} (\bibinfo{year}{2012}), \bibinfo{pages}{441--504}.
\newblock


\bibitem[Kouki et~al\mbox{.}(2017)]%
        {kouki2017user}
\bibfield{author}{\bibinfo{person}{Pigi Kouki}, \bibinfo{person}{James Schaffer}, \bibinfo{person}{Jay Pujara}, \bibinfo{person}{John O'Donovan}, {and} \bibinfo{person}{Lise Getoor}.} \bibinfo{year}{2017}\natexlab{}.
\newblock \showarticletitle{User preferences for hybrid explanations}. In \bibinfo{booktitle}{\emph{Proceedings of the Eleventh ACM Conference on Recommender Systems}}. \bibinfo{pages}{84--88}.
\newblock


\bibitem[Kouki et~al\mbox{.}(2019)]%
        {kouki2019personalized}
\bibfield{author}{\bibinfo{person}{Pigi Kouki}, \bibinfo{person}{James Schaffer}, \bibinfo{person}{Jay Pujara}, \bibinfo{person}{John O'Donovan}, {and} \bibinfo{person}{Lise Getoor}.} \bibinfo{year}{2019}\natexlab{}.
\newblock \showarticletitle{Personalized explanations for hybrid recommender systems}. In \bibinfo{booktitle}{\emph{Proceedings of the 24th international conference on intelligent user interfaces}}. \bibinfo{pages}{379--390}.
\newblock


\bibitem[Kulesza et~al\mbox{.}(2015)]%
        {kulesza2015principles}
\bibfield{author}{\bibinfo{person}{Todd Kulesza}, \bibinfo{person}{Margaret Burnett}, \bibinfo{person}{Weng-Keen Wong}, {and} \bibinfo{person}{Simone Stumpf}.} \bibinfo{year}{2015}\natexlab{}.
\newblock \showarticletitle{Principles of explanatory debugging to personalize interactive machine learning}. In \bibinfo{booktitle}{\emph{Proceedings of the 20th international conference on intelligent user interfaces}}. \bibinfo{pages}{126--137}.
\newblock


\bibitem[Kulesza et~al\mbox{.}(2013)]%
        {kulesza2013too}
\bibfield{author}{\bibinfo{person}{Todd Kulesza}, \bibinfo{person}{Simone Stumpf}, \bibinfo{person}{Margaret Burnett}, \bibinfo{person}{Sherry Yang}, \bibinfo{person}{Irwin Kwan}, {and} \bibinfo{person}{Weng-Keen Wong}.} \bibinfo{year}{2013}\natexlab{}.
\newblock \showarticletitle{Too much, too little, or just right? Ways explanations impact end users' mental models}. In \bibinfo{booktitle}{\emph{2013 IEEE Symposium on visual languages and human centric computing}}. IEEE, \bibinfo{pages}{3--10}.
\newblock


\bibitem[Kunkel et~al\mbox{.}(2019)]%
        {kunkel2019let}
\bibfield{author}{\bibinfo{person}{Johannes Kunkel}, \bibinfo{person}{Tim Donkers}, \bibinfo{person}{Lisa Michael}, \bibinfo{person}{Catalin-Mihai Barbu}, {and} \bibinfo{person}{J{\"u}rgen Ziegler}.} \bibinfo{year}{2019}\natexlab{}.
\newblock \showarticletitle{Let me explain: Impact of personal and impersonal explanations on trust in recommender systems}. In \bibinfo{booktitle}{\emph{Proceedings of the 2019 CHI conference on human factors in computing systems}}. \bibinfo{pages}{1--12}.
\newblock


\bibitem[Kunkel et~al\mbox{.}(2021)]%
        {kunkel2021identifying}
\bibfield{author}{\bibinfo{person}{Johannes Kunkel}, \bibinfo{person}{Thao Ngo}, \bibinfo{person}{J{\"u}rgen Ziegler}, {and} \bibinfo{person}{Nicole Kr{\"a}mer}.} \bibinfo{year}{2021}\natexlab{}.
\newblock \showarticletitle{Identifying group-specific mental models of recommender systems: A novel quantitative approach}. In \bibinfo{booktitle}{\emph{IFIP Conference on Human-Computer Interaction}}. Springer, \bibinfo{pages}{383--404}.
\newblock


\bibitem[Liao and Sundar(2022)]%
        {liao2022designing}
\bibfield{author}{\bibinfo{person}{Q~Vera Liao} {and} \bibinfo{person}{S~Shyam Sundar}.} \bibinfo{year}{2022}\natexlab{}.
\newblock \showarticletitle{Designing for responsible trust in AI systems: A communication perspective}. In \bibinfo{booktitle}{\emph{Proceedings of the 2022 ACM conference on fairness, accountability, and transparency}}. \bibinfo{pages}{1257--1268}.
\newblock


\bibitem[Lim and Dey(2009)]%
        {lim2009assessing}
\bibfield{author}{\bibinfo{person}{Brian~Y Lim} {and} \bibinfo{person}{Anind~K Dey}.} \bibinfo{year}{2009}\natexlab{}.
\newblock \showarticletitle{Assessing demand for intelligibility in context-aware applications}. In \bibinfo{booktitle}{\emph{Proceedings of the 11th international conference on Ubiquitous computing}}. \bibinfo{pages}{195--204}.
\newblock


\bibitem[Lin et~al\mbox{.}(2019)]%
        {lin2019explainable}
\bibfield{author}{\bibinfo{person}{Yujie Lin}, \bibinfo{person}{Pengjie Ren}, \bibinfo{person}{Zhumin Chen}, \bibinfo{person}{Zhaochun Ren}, \bibinfo{person}{Jun Ma}, {and} \bibinfo{person}{Maarten De~Rijke}.} \bibinfo{year}{2019}\natexlab{}.
\newblock \showarticletitle{Explainable outfit recommendation with joint outfit matching and comment generation}.
\newblock \bibinfo{journal}{\emph{IEEE Transactions on Knowledge and Data Engineering}} \bibinfo{volume}{32}, \bibinfo{number}{8} (\bibinfo{year}{2019}), \bibinfo{pages}{1502--1516}.
\newblock


\bibitem[Lins~de Holanda~Coelho et~al\mbox{.}(2020)]%
        {lins2020very}
\bibfield{author}{\bibinfo{person}{Gabriel Lins~de Holanda~Coelho}, \bibinfo{person}{Paul HP~Hanel}, {and} \bibinfo{person}{Lukas J.~Wolf}.} \bibinfo{year}{2020}\natexlab{}.
\newblock \showarticletitle{The very efficient assessment of need for cognition: Developing a six-item version}.
\newblock \bibinfo{journal}{\emph{Assessment}} \bibinfo{volume}{27}, \bibinfo{number}{8} (\bibinfo{year}{2020}), \bibinfo{pages}{1870--1885}.
\newblock


\bibitem[Lu et~al\mbox{.}(2023)]%
        {lu2023user}
\bibfield{author}{\bibinfo{person}{Hongyu Lu}, \bibinfo{person}{Weizhi Ma}, \bibinfo{person}{Yifan Wang}, \bibinfo{person}{Min Zhang}, \bibinfo{person}{Xiang Wang}, \bibinfo{person}{Yiqun Liu}, \bibinfo{person}{Tat-Seng Chua}, {and} \bibinfo{person}{Shaoping Ma}.} \bibinfo{year}{2023}\natexlab{}.
\newblock \showarticletitle{User perception of recommendation explanation: Are your explanations what users need?}
\newblock \bibinfo{journal}{\emph{ACM Transactions on Information Systems}} \bibinfo{volume}{41}, \bibinfo{number}{2} (\bibinfo{year}{2023}), \bibinfo{pages}{1--31}.
\newblock


\bibitem[Madsen and Gregor(2000)]%
        {madsen2000measuring}
\bibfield{author}{\bibinfo{person}{Maria Madsen} {and} \bibinfo{person}{Shirley Gregor}.} \bibinfo{year}{2000}\natexlab{}.
\newblock \showarticletitle{Measuring human-computer trust}. In \bibinfo{booktitle}{\emph{11th australasian conference on information systems}}, Vol.~\bibinfo{volume}{53}. Citeseer, \bibinfo{pages}{6--8}.
\newblock


\bibitem[Martijn et~al\mbox{.}(2022)]%
        {martijn2022knowing}
\bibfield{author}{\bibinfo{person}{Millecamp Martijn}, \bibinfo{person}{Cristina Conati}, {and} \bibinfo{person}{Katrien Verbert}.} \bibinfo{year}{2022}\natexlab{}.
\newblock \showarticletitle{“Knowing me, knowing you”: personalized explanations for a music recommender system}.
\newblock \bibinfo{journal}{\emph{User Modeling and User-Adapted Interaction}} \bibinfo{volume}{32}, \bibinfo{number}{1} (\bibinfo{year}{2022}), \bibinfo{pages}{215--252}.
\newblock


\bibitem[McKnight et~al\mbox{.}(2009)]%
        {mcknight2009trust}
\bibfield{author}{\bibinfo{person}{Harrison McKnight}, \bibinfo{person}{Michelle Carter}, {and} \bibinfo{person}{Paul Clay}.} \bibinfo{year}{2009}\natexlab{}.
\newblock \showarticletitle{Trust in technology: Development of a set of constructs and measures}.
\newblock  (\bibinfo{year}{2009}).
\newblock


\bibitem[Millecamp et~al\mbox{.}(2019)]%
        {millecamp2019explain}
\bibfield{author}{\bibinfo{person}{Martijn Millecamp}, \bibinfo{person}{Nyi~Nyi Htun}, \bibinfo{person}{Cristina Conati}, {and} \bibinfo{person}{Katrien Verbert}.} \bibinfo{year}{2019}\natexlab{}.
\newblock \showarticletitle{To explain or not to explain: the effects of personal characteristics when explaining music recommendations}. In \bibinfo{booktitle}{\emph{Proceedings of the 24th international conference on intelligent user interfaces}}. \bibinfo{pages}{397--407}.
\newblock


\bibitem[Millecamp et~al\mbox{.}(2020)]%
        {millecamp2020s}
\bibfield{author}{\bibinfo{person}{Martijn Millecamp}, \bibinfo{person}{Nyi~Nyi Htun}, \bibinfo{person}{Cristina Conati}, {and} \bibinfo{person}{Katrien Verbert}.} \bibinfo{year}{2020}\natexlab{}.
\newblock \showarticletitle{What's in a user? Towards personalising transparency for music recommender interfaces}. In \bibinfo{booktitle}{\emph{Proceedings of the 28th ACM Conference on User Modeling, Adaptation and Personalization}}. \bibinfo{pages}{173--182}.
\newblock


\bibitem[Miller(2019)]%
        {miller2019explanation}
\bibfield{author}{\bibinfo{person}{Tim Miller}.} \bibinfo{year}{2019}\natexlab{}.
\newblock \showarticletitle{Explanation in artificial intelligence: Insights from the social sciences}.
\newblock \bibinfo{journal}{\emph{Artificial intelligence}}  \bibinfo{volume}{267} (\bibinfo{year}{2019}), \bibinfo{pages}{1--38}.
\newblock


\bibitem[Miller(2022)]%
        {miller2022we}
\bibfield{author}{\bibinfo{person}{Tim Miller}.} \bibinfo{year}{2022}\natexlab{}.
\newblock \showarticletitle{Are we measuring trust correctly in explainability, interpretability, and transparency research?}
\newblock \bibinfo{journal}{\emph{arXiv preprint arXiv:2209.00651}} (\bibinfo{year}{2022}).
\newblock


\bibitem[Munzner(2025)]%
        {munzner2025visualization}
\bibfield{author}{\bibinfo{person}{Tamara Munzner}.} \bibinfo{year}{2025}\natexlab{}.
\newblock \showarticletitle{Visualization analysis and design}. In \bibinfo{booktitle}{\emph{Proceedings of the Special Interest Group on Computer Graphics and Interactive Techniques Conference Courses}}. \bibinfo{pages}{1--2}.
\newblock


\bibitem[Musto et~al\mbox{.}(2019)]%
        {musto2019justifying}
\bibfield{author}{\bibinfo{person}{Cataldo Musto}, \bibinfo{person}{Pasquale Lops}, \bibinfo{person}{Marco de Gemmis}, {and} \bibinfo{person}{Giovanni Semeraro}.} \bibinfo{year}{2019}\natexlab{}.
\newblock \showarticletitle{Justifying recommendations through aspect-based sentiment analysis of users reviews}. In \bibinfo{booktitle}{\emph{Proceedings of the 27th ACM conference on user modeling, adaptation and personalization}}. \bibinfo{pages}{4--12}.
\newblock


\bibitem[Musto et~al\mbox{.}(2016)]%
        {musto2016explod}
\bibfield{author}{\bibinfo{person}{Cataldo Musto}, \bibinfo{person}{Fedelucio Narducci}, \bibinfo{person}{Pasquale Lops}, \bibinfo{person}{Marco De~Gemmis}, {and} \bibinfo{person}{Giovanni Semeraro}.} \bibinfo{year}{2016}\natexlab{}.
\newblock \showarticletitle{Explod: A framework for explaining recommendations based on the linked open data cloud}. In \bibinfo{booktitle}{\emph{Proceedings of the 10th ACM Conference on Recommender Systems}}. \bibinfo{pages}{151--154}.
\newblock


\bibitem[Naveed et~al\mbox{.}(2018)]%
        {naveed2018argumentation}
\bibfield{author}{\bibinfo{person}{Sidra Naveed}, \bibinfo{person}{Tim Donkers}, {and} \bibinfo{person}{J{\"u}rgen Ziegler}.} \bibinfo{year}{2018}\natexlab{}.
\newblock \showarticletitle{Argumentation-based explanations in recommender systems: Conceptual framework and empirical results}. In \bibinfo{booktitle}{\emph{Adjunct Publication of the 26th Conference on User Modeling, Adaptation and Personalization}}. \bibinfo{pages}{293--298}.
\newblock


\bibitem[Nunes and Jannach(2017)]%
        {nunes2017systematic}
\bibfield{author}{\bibinfo{person}{Ingrid Nunes} {and} \bibinfo{person}{Dietmar Jannach}.} \bibinfo{year}{2017}\natexlab{}.
\newblock \showarticletitle{A systematic review and taxonomy of explanations in decision support and recommender systems}.
\newblock \bibinfo{journal}{\emph{User Modeling and User-Adapted Interaction}} \bibinfo{volume}{27}, \bibinfo{number}{3} (\bibinfo{year}{2017}), \bibinfo{pages}{393--444}.
\newblock


\bibitem[Ooge et~al\mbox{.}(2022)]%
        {ooge2022explaining}
\bibfield{author}{\bibinfo{person}{Jeroen Ooge}, \bibinfo{person}{Shotallo Kato}, {and} \bibinfo{person}{Katrien Verbert}.} \bibinfo{year}{2022}\natexlab{}.
\newblock \showarticletitle{Explaining recommendations in e-learning: Effects on adolescents' trust}. In \bibinfo{booktitle}{\emph{Proceedings of the 27th International Conference on Intelligent User Interfaces}}. \bibinfo{pages}{93--105}.
\newblock


\bibitem[Pu et~al\mbox{.}(2011)]%
        {pu2011user}
\bibfield{author}{\bibinfo{person}{Pearl Pu}, \bibinfo{person}{Li Chen}, {and} \bibinfo{person}{Rong Hu}.} \bibinfo{year}{2011}\natexlab{}.
\newblock \showarticletitle{A user-centric evaluation framework for recommender systems}. In \bibinfo{booktitle}{\emph{Proceedings of the fifth ACM conference on Recommender systems}}. \bibinfo{pages}{157--164}.
\newblock


\bibitem[Pu et~al\mbox{.}(2012)]%
        {pu2012evaluating}
\bibfield{author}{\bibinfo{person}{Pearl Pu}, \bibinfo{person}{Li Chen}, {and} \bibinfo{person}{Rong Hu}.} \bibinfo{year}{2012}\natexlab{}.
\newblock \showarticletitle{Evaluating recommender systems from the user’s perspective: survey of the state of the art}.
\newblock \bibinfo{journal}{\emph{User Modeling and User-Adapted Interaction}} \bibinfo{volume}{22}, \bibinfo{number}{4} (\bibinfo{year}{2012}), \bibinfo{pages}{317--355}.
\newblock


\bibitem[Sato et~al\mbox{.}(2019)]%
        {sato2019context}
\bibfield{author}{\bibinfo{person}{Masahiro Sato}, \bibinfo{person}{Koki Nagatani}, \bibinfo{person}{Takashi Sonoda}, \bibinfo{person}{Qian Zhang}, {and} \bibinfo{person}{Tomoko Ohkuma}.} \bibinfo{year}{2019}\natexlab{}.
\newblock \showarticletitle{Context style explanation for recommender systems}.
\newblock \bibinfo{journal}{\emph{Journal of Information Processing}}  \bibinfo{volume}{27} (\bibinfo{year}{2019}), \bibinfo{pages}{720--729}.
\newblock


\bibitem[Siepmann and Chatti(2023)]%
        {siepmann2023trust}
\bibfield{author}{\bibinfo{person}{Clara Siepmann} {and} \bibinfo{person}{Mohamed~Amine Chatti}.} \bibinfo{year}{2023}\natexlab{}.
\newblock \showarticletitle{Trust and transparency in recommender systems}.
\newblock \bibinfo{journal}{\emph{arXiv preprint arXiv:2304.08094}} (\bibinfo{year}{2023}).
\newblock


\bibitem[Sun et~al\mbox{.}(2021)]%
        {sun2021unsupervised}
\bibfield{author}{\bibinfo{person}{Peijie Sun}, \bibinfo{person}{Le Wu}, \bibinfo{person}{Kun Zhang}, \bibinfo{person}{Yu Su}, {and} \bibinfo{person}{Meng Wang}.} \bibinfo{year}{2021}\natexlab{}.
\newblock \showarticletitle{An unsupervised aspect-aware recommendation model with explanation text generation}.
\newblock \bibinfo{journal}{\emph{ACM Transactions on Information Systems (TOIS)}} \bibinfo{volume}{40}, \bibinfo{number}{3} (\bibinfo{year}{2021}), \bibinfo{pages}{1--29}.
\newblock


\bibitem[Szymanski et~al\mbox{.}(2021)]%
        {millecamp2021textualvisual}
\bibfield{author}{\bibinfo{person}{Maxwell Szymanski}, \bibinfo{person}{Martijn Millecamp}, {and} \bibinfo{person}{Katrien Verbert}.} \bibinfo{year}{2021}\natexlab{}.
\newblock \showarticletitle{Visual, textual or hybrid: the effect of user expertise on different explanations}. In \bibinfo{booktitle}{\emph{Proceedings of the 26th international conference on intelligent user interfaces}}. \bibinfo{pages}{109--119}.
\newblock


\bibitem[Tintarev(2017)]%
        {tintarev2017presenting}
\bibfield{author}{\bibinfo{person}{Nava Tintarev}.} \bibinfo{year}{2017}\natexlab{}.
\newblock \showarticletitle{Presenting diversity aware recommendations}. In \bibinfo{booktitle}{\emph{FATREC Workshop on Responsible Recommendation Proceedings}}.
\newblock


\bibitem[Tintarev and Masthoff(2007)]%
        {tintarev2007survey}
\bibfield{author}{\bibinfo{person}{Nava Tintarev} {and} \bibinfo{person}{Judith Masthoff}.} \bibinfo{year}{2007}\natexlab{}.
\newblock \showarticletitle{A survey of explanations in recommender systems}. In \bibinfo{booktitle}{\emph{2007 IEEE 23rd international conference on data engineering workshop}}. IEEE, \bibinfo{pages}{801--810}.
\newblock


\bibitem[Tintarev and Masthoff(2010)]%
        {tintarev2010designing}
\bibfield{author}{\bibinfo{person}{Nava Tintarev} {and} \bibinfo{person}{Judith Masthoff}.} \bibinfo{year}{2010}\natexlab{}.
\newblock \showarticletitle{Designing and evaluating explanations for recommender systems}.
\newblock In \bibinfo{booktitle}{\emph{Recommender systems handbook}}. \bibinfo{publisher}{Springer}, \bibinfo{pages}{479--510}.
\newblock


\bibitem[Tintarev and Masthoff(2015)]%
        {Tintarev2015}
\bibfield{author}{\bibinfo{person}{Nava Tintarev} {and} \bibinfo{person}{Judith Masthoff}.} \bibinfo{year}{2015}\natexlab{}.
\newblock \bibinfo{booktitle}{\emph{Explaining Recommendations: Design and Evaluation}}.
\newblock \bibinfo{publisher}{Springer US}, \bibinfo{address}{Boston, MA}, \bibinfo{pages}{353--382}.
\newblock
\showISBNx{978-1-4899-7637-6}
\href{https://doi.org/10.1007/978-1-4899-7637-6_10}{doi:\nolinkurl{10.1007/978-1-4899-7637-6_10}}


\bibitem[Tsai and Brusilovsky(2019)]%
        {tsai2019explaining}
\bibfield{author}{\bibinfo{person}{Chun-Hua Tsai} {and} \bibinfo{person}{Peter Brusilovsky}.} \bibinfo{year}{2019}\natexlab{}.
\newblock \showarticletitle{Explaining recommendations in an interactive hybrid social recommender}. In \bibinfo{booktitle}{\emph{Proceedings of the 24th international conference on intelligent user interfaces}}. \bibinfo{pages}{391--396}.
\newblock


\bibitem[Vig et~al\mbox{.}(2009)]%
        {vig2009tagsplanations}
\bibfield{author}{\bibinfo{person}{Jesse Vig}, \bibinfo{person}{Shilad Sen}, {and} \bibinfo{person}{John Riedl}.} \bibinfo{year}{2009}\natexlab{}.
\newblock \showarticletitle{Tagsplanations: explaining recommendations using tags}. In \bibinfo{booktitle}{\emph{Proceedings of the 14th international conference on Intelligent user interfaces}}. \bibinfo{pages}{47--56}.
\newblock


\bibitem[Wang et~al\mbox{.}(2019)]%
        {wang2019designing}
\bibfield{author}{\bibinfo{person}{Danding Wang}, \bibinfo{person}{Qian Yang}, \bibinfo{person}{Ashraf Abdul}, {and} \bibinfo{person}{Brian~Y Lim}.} \bibinfo{year}{2019}\natexlab{}.
\newblock \showarticletitle{Designing theory-driven user-centric explainable AI}. In \bibinfo{booktitle}{\emph{Proceedings of the 2019 CHI conference on human factors in computing systems}}. \bibinfo{pages}{1--15}.
\newblock


\bibitem[Ware(2019)]%
        {ware2019information}
\bibfield{author}{\bibinfo{person}{Colin Ware}.} \bibinfo{year}{2019}\natexlab{}.
\newblock \bibinfo{booktitle}{\emph{Information visualization: perception for design}}.
\newblock \bibinfo{publisher}{Morgan Kaufmann}.
\newblock


\bibitem[Xian et~al\mbox{.}(2020)]%
        {xian2020cafe}
\bibfield{author}{\bibinfo{person}{Yikun Xian}, \bibinfo{person}{Zuohui Fu}, \bibinfo{person}{Handong Zhao}, \bibinfo{person}{Yingqiang Ge}, \bibinfo{person}{Xu Chen}, \bibinfo{person}{Qiaoying Huang}, \bibinfo{person}{Shijie Geng}, \bibinfo{person}{Zhou Qin}, \bibinfo{person}{Gerard De~Melo}, \bibinfo{person}{Shan Muthukrishnan}, {et~al\mbox{.}}} \bibinfo{year}{2020}\natexlab{}.
\newblock \showarticletitle{CAFE: Coarse-to-fine neural symbolic reasoning for explainable recommendation}. In \bibinfo{booktitle}{\emph{Proceedings of the 29th ACM International Conference on Information \& Knowledge Management}}. \bibinfo{pages}{1645--1654}.
\newblock


\bibitem[Yang et~al\mbox{.}(2020)]%
        {yang2020visual}
\bibfield{author}{\bibinfo{person}{Fumeng Yang}, \bibinfo{person}{Zhuanyi Huang}, \bibinfo{person}{Jean Scholtz}, {and} \bibinfo{person}{Dustin~L Arendt}.} \bibinfo{year}{2020}\natexlab{}.
\newblock \showarticletitle{How do visual explanations foster end users' appropriate trust in machine learning?}. In \bibinfo{booktitle}{\emph{Proceedings of the 25th international conference on intelligent user interfaces}}. \bibinfo{pages}{189--201}.
\newblock


\bibitem[Yang et~al\mbox{.}(2024)]%
        {yang2024fine}
\bibfield{author}{\bibinfo{person}{Mengyuan Yang}, \bibinfo{person}{Mengying Zhu}, \bibinfo{person}{Yan Wang}, \bibinfo{person}{Linxun Chen}, \bibinfo{person}{Yilei Zhao}, \bibinfo{person}{Xiuyuan Wang}, \bibinfo{person}{Bing Han}, \bibinfo{person}{Xiaolin Zheng}, {and} \bibinfo{person}{Jianwei Yin}.} \bibinfo{year}{2024}\natexlab{}.
\newblock \showarticletitle{Fine-tuning large language model based explainable recommendation with explainable quality reward}. In \bibinfo{booktitle}{\emph{Proceedings of the AAAI Conference on Artificial Intelligence}}, Vol.~\bibinfo{volume}{38}. \bibinfo{pages}{9250--9259}.
\newblock


\bibitem[Zhang et~al\mbox{.}(2020)]%
        {zhang2020explainable}
\bibfield{author}{\bibinfo{person}{Yongfeng Zhang}, \bibinfo{person}{Xu Chen}, {et~al\mbox{.}}} \bibinfo{year}{2020}\natexlab{}.
\newblock \showarticletitle{Explainable recommendation: A survey and new perspectives}.
\newblock \bibinfo{journal}{\emph{Foundations and Trends{\textregistered} in Information Retrieval}} \bibinfo{volume}{14}, \bibinfo{number}{1} (\bibinfo{year}{2020}), \bibinfo{pages}{1--101}.
\newblock


\bibitem[Zhao et~al\mbox{.}(2019b)]%
        {zhao2019personalized}
\bibfield{author}{\bibinfo{person}{Guoshuai Zhao}, \bibinfo{person}{Hao Fu}, \bibinfo{person}{Ruihua Song}, \bibinfo{person}{Tetsuya Sakai}, \bibinfo{person}{Zhongxia Chen}, \bibinfo{person}{Xing Xie}, {and} \bibinfo{person}{Xueming Qian}.} \bibinfo{year}{2019}\natexlab{b}.
\newblock \showarticletitle{Personalized reason generation for explainable song recommendation}.
\newblock \bibinfo{journal}{\emph{ACM Transactions on Intelligent Systems and Technology (TIST)}} \bibinfo{volume}{10}, \bibinfo{number}{4} (\bibinfo{year}{2019}), \bibinfo{pages}{1--21}.
\newblock


\bibitem[Zhao et~al\mbox{.}(2018)]%
        {zhao2018you}
\bibfield{author}{\bibinfo{person}{Guoshuai Zhao}, \bibinfo{person}{Hao Fu}, \bibinfo{person}{Ruihua Song}, \bibinfo{person}{Tetsuya Sakai}, \bibinfo{person}{Xing Xie}, {and} \bibinfo{person}{Xueming Qian}.} \bibinfo{year}{2018}\natexlab{}.
\newblock \showarticletitle{Why you should listen to this song: Reason generation for explainable recommendation}. In \bibinfo{booktitle}{\emph{2018 IEEE International Conference on Data Mining Workshops (ICDMW)}}. IEEE, \bibinfo{pages}{1316--1322}.
\newblock


\bibitem[Zhao et~al\mbox{.}(2019a)]%
        {zhao2019users}
\bibfield{author}{\bibinfo{person}{Ruijing Zhao}, \bibinfo{person}{Izak Benbasat}, {and} \bibinfo{person}{Hasan Cavusoglu}.} \bibinfo{year}{2019}\natexlab{a}.
\newblock \showarticletitle{Do users always want to know more? Investigating the relationship between system transparency and users’ trust in advice-giving systems}. In \bibinfo{booktitle}{\emph{Proceedings of the 27th European Conference on Information Systems}}.
\newblock


\end{thebibliography}
\end{document}